\documentclass[11pt,nofootinbib,preprintnumbers,superscriptaddress]{revtex4-1}

\usepackage{amssymb, amsfonts, amsmath, mathtools, bm}
\usepackage{color}

\usepackage{mathrsfs}
\usepackage{enumerate}
\usepackage{amsfonts}
\usepackage{graphicx}
\newcommand{\bra}[1]{\left( #1 \right)}
\newcommand{\brb}[1]{\left[ #1 \right]}
\newcommand{\brc}[1]{\left\{ #1 \right\}}
\newcommand{\norm}[1]{\left| #1 \right|}

\usepackage[
]{hyperref}
\hypersetup{%
 setpagesize=false,
 bookmarksnumbered=true,%
 bookmarksopen=true,%
 colorlinks=true,%
 linkcolor=blue,%
 citecolor=red}

\begin{document}
\title{
{\Large
Black hole ringdown from physically sensible initial value problem in higher-order scalar-tensor theories
}}
\author{Keisuke Nakashi}
\affiliation{Department of Social Design Engineering, National Institute of Technology (KOSEN), Kochi College, 200-1 Monobe Otsu, Nankoku, Kochi, 783-8508, Japan}
\affiliation{Department of Physics, Rikkyo University, Toshima, Tokyo 171-8501, Japan}
\author{Masashi Kimura}
\affiliation{Department of Informatics and Electronics, Daiichi Institute of Technology, Tokyo 110-0005, Japan}
\affiliation{Department of Physics, Rikkyo University, Toshima, Tokyo 171-8501, Japan}
\author{Hayato Motohashi}
\affiliation{Division of Liberal Arts, Kogakuin University, 2665-1 Nakano-machi, Hachioji, Tokyo 192-0015, Japan}
\author{Kazufumi Takahashi}
\affiliation{Center for Gravitational Physics and Quantum Information, Yukawa Institute for Theoretical Physics, Kyoto University, 606-8502, Kyoto, Japan}
\date{\today}
\preprint{RUP-23-19, YITP-23-118}

\begin{abstract}
  We study odd-parity perturbations about static and spherically symmetric black hole solutions with a linearly time-dependent scalar field in higher-order scalar-tensor theories. 
  In particular, we consider stealth Schwarzschild and stealth Schwarzschild-de Sitter solutions, where the deviation from the general relativity case is controlled by a single parameter. 
  We find that complex frequencies of quasinormal modes (QNMs) are given by a simple scaling of those in general relativity.  
  We also show that there is a degeneracy between the parameter characterizing the modification from general relativity and the black hole mass. 
  We then consider a physically sensible initial value problem by taking into account the fact that the effective metric for the odd-parity perturbations is in general different from the background metric.
  We confirm that damped oscillations appearing at late times are indeed dominated by the QNMs. 
  Our analysis includes the case where the perturbations are superluminal, and we demonstrate in this case that the perturbations can escape from the region inside the horizon for the background metric.
\end{abstract}

\maketitle

\section{Introduction}\label{sec:intro}

Testing gravity has been a central issue in physics.
Apart from cosmological tests of gravity~\cite{Koyama:2015vza,Ferreira:2019xrr,Arai:2022ilw}, there have been an increasing number of gravitational-wave events from binary black hole mergers, which offer a possibility to test gravity at strong-field/dynamical regimes.
In general relativity (GR), the late-time gravitational wave signal emitted from binary black hole mergers, known as the ringdown signal, can be well described by a superposition of quasinormal modes (QNMs)~\cite{Buonanno:2006ui}.
Each QNM is characterized by a specific complex frequency, whose real and imaginary parts respectively correspond to the frequency of temporal oscillation and the exponential damping rate.
The no-hair theorem of black holes in (vacuum) GR implies that the QNM frequencies are determined solely by the mass and angular momentum of the black hole.
However, in modified gravity, black holes can support some nontrivial hair other than the mass and angular momentum, which would affect the QNM spectrum.
In other words, the information about the underlying gravitational theory would be encoded in the QNM spectrum.

In contrast to GR where gravity is described solely by the spacetime metric, modified gravity theories in general involve additional degrees of freedom.
The simplest class of modified gravity is the class of scalar-tensor theories, where a single scalar field represents the modification of gravity.
Starting with the seminal theory of Brans-Dicke~\cite{Brans:1961sx}, a number of scalar-tensor theories have been proposed so far.
Horndeski theories~\cite{Horndeski:1974wa,Deffayet:2011gz,Kobayashi:2011nu}, which form the most general class of scalar-tensor theories with second-order Euler-Lagrange equations, provide a unified description of such traditional theories.
It should be noted that the second-order nature of the Euler-Lagrange equations guarantees the absence of the Ostrogradsky ghost~\cite{Woodard:2015zca,Motohashi:2014opa,Motohashi:2020psc,Aoki:2020gfv}.

Meanwhile, the Horndeski class is not the most general class of ghost-free scalar-tensor theories.
Indeed, even if the Euler-Lagrange equations contain higher-order derivatives, the problem of Ostrogradsky ghost can be circumvented by imposing the degeneracy condition~\cite{Motohashi:2014opa,Langlois:2015cwa,Motohashi:2016ftl,Klein:2016aiq,Motohashi:2017eya,Motohashi:2018pxg}.
Extensions of Horndeski theories in this direction are called degenerate higher-order scalar-tensor (DHOST) theories~\cite{Langlois:2015cwa,Crisostomi:2016czh,BenAchour:2016fzp}.
Another systematic way to extend the Horndeski class is to employ the disformal transformation~\cite{Bekenstein:1992pj,Bruneton:2007si,Bettoni:2013diz} and its generalization involving higher derivatives of the scalar field~\cite{Takahashi:2021ttd,Takahashi:2023vva}.
In fact, the disformal transformation maps the Horndeski class to (a particular subclass of) the DHOST class, while the generalized disformal transformation yields a larger class of ghost-free theories, which is called the generalized disformal Horndeski (GDH) class~\cite{Takahashi:2022mew}.\footnote{Matter coupling could introduce an Ostrogradsky mode in generalized disformal Horndeski theories in general, while there exists a nontrivial subclass where this problem can be avoided~\cite{Takahashi:2022mew,Naruko:2022vuh,Takahashi:2022ctx,Ikeda:2023ntu}.}
A yet further extension can be obtained by relaxing the degeneracy condition in such a way that it is satisfied only under the unitary gauge.
Away from the unitary gauge, apparently there is an Ostrogradsky mode, but it actually satisfies an elliptic differential equation on a spacelike hypersurface and hence does not propagate.
Such a mode is often called a shadowy mode~\cite{DeFelice:2018ewo,DeFelice:2021hps}, which itself is harmless.
By allowing for the existence of the shadowy mode, one obtains U-DHOST~\cite{DeFelice:2018ewo,DeFelice:2021hps,DeFelice:2022xvq} and generalized disformal unitary-degenerate (GDU) theories~\cite{Takahashi:2023jro}.

An interesting class of solutions in scalar-tensor theories is the so-called stealth solution, where the metric is the same as in a GR solution but the scalar field has a nontrivial profile.
The stealth solutions have been found and studied in the Brans-Dicke theory~\cite{Nariai1968,OHanlon:1972ysn,BARROW1990294,Romero1993,
Kolitch:1994kr,Johri:1994rw,Giardino:2022sdv,Giardino:2023qlu}, more general scalar-tensor theories~\cite{Ayon-Beato:2004nzi,Ayon-Beato:2005yoq,Mukohyama:2005rw,Robinson:2006ib,Ayon-Beato:2015qfa,Alvarez:2016qky,Smolic:2017bic,Franzin:2021yvf}, and Horndeski and DHOST theories~\cite{Babichev:2013cya,Kobayashi:2014eva,Babichev:2016kdt,Babichev:2017lmw,Minamitsuji:2018vuw,BenAchour:2018dap,Motohashi:2018wdq,Motohashi:2019sen,Minamitsuji:2019shy,Bernardo:2019yxp,Charmousis:2019vnf,Takahashi:2020hso,Bernardo:2020ehy,Gorji:2020bfl}.
In particular, the general construction of stealth solutions was developed in \cite{Motohashi:2018wdq,Takahashi:2020hso} in a covariant manner. 
The perturbation theory about stealth black hole solutions has been studied extensively~\cite{Babichev:2018uiw,Takahashi:2019oxz,deRham:2019gha,Motohashi:2019ymr,Khoury:2020aya,Tomikawa:2021pca,Takahashi:2021bml,Mukohyama:2022skk,Khoury:2022zor}.
It then turned out that perturbations of stealth solutions are strongly coupled in DHOST theories~\cite{Babichev:2018uiw,deRham:2019gha,Motohashi:2019ymr,Takahashi:2021bml}, and this problem is expected to persist in GDH theories.
A possible way out of this problem is to consider a small detuning (i.e., scordatura) of the degeneracy condition~\cite{Motohashi:2019ymr}.\footnote{The scordatura term affects the stealth black hole background, leading to a time-dependent correction. However, the time dependence is typically very weak and can be negligible at astrophysical scales~\cite{Mukohyama:2005rw,DeFelice:2022qaz}.}
This would introduce an Ostrogradsky mode in general, but its mass can be pushed above the cutoff of the theory.
Moreover, it is even possible to have the scordatura term in U-DHOST theories that are intrinsically free of Ostrogradsky ghost~\cite{DeFelice:2022xvq}.
Therefore, DHOST (or GDH) theories supplemented with the scordatura term would provide a consistent description of stealth solutions.

In the present paper, we perform a time-domain analysis of perturbations about stealth black hole solutions in DHOST theories.
In doing so, the main difficulty comes from the fact that the effective metric (i.e., the one on which the perturbations propagate) is in general different from the background metric which determines the motion of (minimally coupled) matter fields. 
This implies that a portion of a hypersurface which is spacelike with respect to the effective metric can be timelike with respect to the background metric.
Therefore, when matter fields are taken into account, one has to carefully choose the initial hypersurface so that it is spacelike with respect to both the effective metric and the background metric.
This issue has been addressed in \cite{Nakashi:2022wdg} for the case of monopole perturbations about stealth black hole solutions in DHOST theories.
The aim of the present paper is to extend the analysis of \cite{Nakashi:2022wdg} to odd-parity perturbations.

The rest of this paper is organized as follows. 
In Sec.~\ref{sec:gravityback}, we explain the DHOST theories and their stealth black hole solutions.
In addition, following~\cite{Takahashi:2019oxz}, we analyze the odd-parity perturbations about the stealth black hole solutions to see that one has to introduce a new time coordinate (called $\tilde{t}$) to recast the master equation for the odd-parity perturbations in the form of a wave equation.
In Sec.~\ref{sec:Schwarzschildcase}, we discuss the effective metric, the character of a constant-$\tilde{t}$ hypersurface, and characteristic curves for the odd-parity perturbations about the stealth Schwarzschild solutions. 
We also discuss QNM frequencies in the DHOST theories and obtain the time evolution of the perturbations employing the physically sensible formulation of an initial value problem developed in \cite{Nakashi:2022wdg}. 
In particular, we confirm that the numerical waveform exhibits damped oscillations at late times, which can be well fitted by a superposition of the QNMs for the DHOST theories. 
In Sec.~\ref{sec:SdScase}, we perform a similar analysis for the stealth Schwarzschild-de Sitter solutions. 
Finally, we draw our conclusions in Sec.~\ref{sec:summary}. 
In what follows, we use the geometric units in which $c=G=1$.

\section{gravity theory, Background and Odd-parity perturbations}
\label{sec:gravityback}

\subsection{Gravity theory}

The action of the quadratic DHOST theories is given by~\cite{Langlois:2015cwa}
\begin{align}
    S = \int {\rm d}^{4}x \sqrt{-g} 
    \brb{ 
      F_{0}(\phi, X)
    + F_{1}(\phi, X) \Box \phi
    + F_{2}(\phi, X) R
    + \sum_{I=1}^{5} A_{I}(\phi, X) L_{I}^{(2)}
    },
    \label{eq:actionDHOST}
\end{align}
where the coupling functions~$F_{0}, F_{1}, F_{2},$ and $A_{I}$ are functions of the scalar field~$\phi$ and its kinetic term~$X = \phi_{\mu} \phi^{\mu}$ 
and 
\begin{align}
\begin{split}
    &L_{1}^{(2)} = \phi_{\mu \nu}\phi^{\mu \nu},
    \qquad
    L_{2}^{(2)} = (\Box \phi)^{2}, 
    \qquad
    L_{3}^{(2)} = \phi^{\mu} \phi_{\mu \nu} \phi^{\nu} \Box \phi,
    \\
    &L_{4}^{(2)} = \phi^{\mu} \phi_{\mu \nu} \phi^{\nu \lambda} \phi_{\lambda}, 
    \qquad
    L_{5}^{(2)} = (\phi^{\mu} \phi_{\mu \nu} \phi^{\nu})^{2},
\end{split}
\end{align}
with $\phi_{\mu} = \nabla_{\mu} \phi$ and $\phi_{\mu \nu} = \nabla_{\mu} \nabla_{\nu} \phi$. 
For a generic choice of the coupling functions, the theory described by the action~\eqref{eq:actionDHOST} suffers from the problem of the Ostrogradsky ghost associated with higher derivatives in the equations of motion. 
The Ostrogradsky ghost can be removed by imposing the following degeneracy conditions:
\begin{align}
\begin{split}
    A_{2} &= - A_{1} \neq - \frac{F_{2}}{X},
    \\
    A_{4} &= \frac{1}{8(F_{2}-XA_{1})^{2}}
    \left\{
    4F_{2}\brb{3(A_{1}-2F_{2X})^{2} - 2A_{3}F_{2}} - A_{3}X^{2}(16A_{1}F_{2X}+A_{3}F_{2}) \right.
    \\
    &\quad \left.
    +4X(3A_{1}A_{3}F_{2} + 16A_{1}^{2}F_{2X} - 16A_{1}F_{2X}^{2} - 4A_{1}^{3} + 2A_{3}F_{2}F_{2X})
    \right\},
    \\
    A_{5} &= \frac{1}{8(F_{2}-XA_{1})^{2}}(2A_{1}-XA_{3}-4F_{2X}) \brb{A_{1}(2A_{1} + 3XA_{3} - 4F_{2X}) - 4A_{3}F_{2}},
\end{split}
    \label{eq:degeneracycons}
\end{align}
where a subscript~$X$ denotes the derivative with respect to $X$. 
The DHOST theories described by Eq.~\eqref{eq:actionDHOST} with the degeneracy conditions~\eqref{eq:degeneracycons} is called class Ia~\cite{Langlois:2015cwa,BenAchour:2016cay}, which can be mapped to the Horndeski theory via disformal transformation. 
It is known that all the other classes of quadratic DHOST theories are phenomenologically disfavored in the sense that either the cosmological perturbations are unstable or the modes correspond to gravitational waves are absent.

In the present paper, we consider a subclass of the class Ia quadratic DHOST theories, which is described by the following action: 
\begin{align}
    S = \int {\rm d}x^{4} \sqrt{-g}
    \brb{
    F_{0}(X) + F_{2}(X)R
    + \sum_{I=1}^{5} A_{I}(X) L_{I}^{(2)}
    },
    \label{eq:action_subDHOST}
\end{align}
where we have set $F_{1}=0$ and assumed that the coupling functions are functions only of $X$. 
In other words, we focus on the subclass of the quadratic DHOST theories whose action is invariant under the shift ($\phi \to \phi + {\rm const}.$) and the reflection ($\phi \to - \phi$) of the scalar field.
As we will see in the next subsection, these theories admit an interesting class of solutions known as the stealth solutions, i.e., a GR solution with a linearly time-dependent scalar field.

\subsection{Background spacetime and scalar field}\label{subsec:BG}

We consider a static and spherically symmetric background spacetime. 
The metric of the background spacetime is given by
\begin{align}
    \bar{g}_{\mu \nu} {\rm d}x^{\mu} {\rm d}x^{\nu}
    = 
    - A(r) {\rm d}t^{2}
    + \frac{{\rm d}r^{2}}{B(r)} 
    + r^{2} \gamma_{ab} {\rm d}x^{a} {\rm d}x^{b},
    \label{eq:generalmetricBG}
\end{align}
where $\gamma_{ab}$ is the metric on a two-dimensional unit sphere, $\gamma_{ab} {\rm d}x^{a} {\rm d}x^{b} = {\rm d}\theta^{2} + \sin^{2} \theta {\rm d} \varphi^{2}$. 
As for the scalar field, we impose the following ansatz:
\begin{align}
    \bar{\phi}(t,r) = qt + \psi(r),
\end{align}
where $q$ is a nonvanishing constant. 
We note that the linear time dependence of the scalar field is compatible with the static metric because the action~\eqref{eq:action_subDHOST} depends on the scalar field only through its derivatives. 
Having said that, the linear time dependence can be still allowed in theories without shift symmetry~\cite{Minamitsuji:2018vuw,Motohashi:2018wdq,Takahashi:2020hso}.

In the present paper, in particular, we focus on stealth black hole solutions. 
A stealth black hole solution is described by the metric which is the same as the one in GR, while the scalar field has a nontrivial configuration. 
The general construction of stealth solutions was developed in~\cite{Motohashi:2018wdq,Takahashi:2020hso} in a covariant manner. 
The idea is to substitute the metric and scalar field ansatz into the equations of motion and derive the conditions on the coupling functions of DHOST theories under which the equations are trivially satisfied.
Assuming that $X=-q^2$, the stealth Schwarzschild-de Sitter (dS) metric,
    \begin{align}
    A(r) = B(r) = 1 - \frac{r_{\rm s}}{r} - \frac{\Lambda r^{2}}{3},
    \end{align}
with $r_{\rm s}$ and $\Lambda$ being constants, can be a solution if the following conditions are satisfied~\cite{Motohashi:2018wdq,Takahashi:2020hso}:
    \begin{align}
    \left. \brc{  F_{0}+2\Lambda \bra{F_{2} - XA_{1}} } \right|_{X=-q^{2}}=0,\,\,
    \left. \brc{ 2F_{0X}+\Lambda \bra{8F_{2X}-2A_{1} + 4XA_{1X} + 3XA_{3}} } \right|_{X=-q^{2}}=0.
    \label{existence_condition}
    \end{align}
Note that, among the three degeneracy conditions in \eqref{eq:degeneracycons}, we have used only $A_2=-A_1$ in deriving the above conditions.
Therefore, the stealth Schwarzschild-dS solution exists even away from the DHOST theories so long as $A_2=-A_1$. 
Note also that the stealth Schwarzschild solution can be realized by putting $\Lambda=0$.
In this case, the above condition reads
\begin{align}
    \left. F_{0}\right|_{X=-q^{2}} = 0,
    \qquad
    \left. F_{0X}\right|_{X=-q^{2}} = 0.
\end{align}

For the stealth black hole solutions, the scalar field profile can be obtained from the condition~$X=-q^2$ as follows:
\begin{align}
    \bar{\phi} = q \bra{t
     \pm \int \frac{\sqrt{1-A(r)}}{A(r)}{\rm d}r}.
\end{align}
Here, we choose the plus branch so that $\phi$ is regular at the future event horizon. 
Indeed, for the plus branch, the behavior of the scalar field near the future event horizon where $A(r) \simeq0$ can be approximated as
\begin{align}
    \bar{\phi} \simeq q \bra{t + \int \frac{{\rm d}r}{A(r)}} = qv,
\end{align}
where $v$ is the ingoing Eddington-Finkelstein coordinate defined by $v \coloneqq t + \int A(r)^{-1} {\rm d}r$.

\subsection{Odd-parity perturbations: quadratic Lagrangian and equation of motion}
\label{subsec:odd}

We study linear odd-parity perturbations around a static and spherically symmetric spacetime in DHOST theories. 
Although we will focus on the stealth black hole solutions in the subsequent sections, for the time being, we investigate the perturbations around a general static and spherically symmetric spacetime described by the metric~\eqref{eq:generalmetricBG}, following the discussion in \cite{Takahashi:2019oxz,Takahashi:2021bml}. 
To study the odd-parity perturbations, we define the metric perturbation as $\epsilon h_{\mu \nu} \coloneqq g_{\mu \nu} - \bar{g}_{\mu \nu}$, where $\epsilon$ is a small parameter. 
Due to the spherical symmetry of the background spacetime, it is useful to expand the odd-parity perturbations in terms of the spherical harmonics~$Y_{\ell m}(\theta, \varphi)$ as follows: 
\begin{align}
\begin{split}
    h_{tt} &= h_{tr} = h_{rr} = 0,
    \\
    h_{ta} &= \sum_{\ell,m} h_{0,\ell m}(t,r) E_{a}{}^{b} \bar \nabla_{b} Y_{\ell m}(\theta, \varphi),
    \\
    h_{ra} &= \sum_{\ell,m}h_{1,\ell m}(t,r) E_{a}{}^{b} \bar \nabla_{b} Y_{\ell m}(\theta,\varphi),
    \\
    h_{ab} &= 
    \sum_{\ell,m} h_{2,\ell m}(t,r) E_{(a}{}^{c} \bar \nabla_{b)} \bar \nabla_{c}Y_{\ell m}(\theta,\varphi),
\end{split}
\end{align}
where $E_{ab}$ is the completely antisymmetric tensor defined on a two-dimensional unit sphere, and $\bar \nabla_{a}$ denotes the covariant derivative with respect to $\gamma_{ab}$. 
Due to the symmetry of the background spacetime, it is sufficient to consider only $m=0$. 
We note that the odd-parity perturbations do not have $\ell = 0$ mode, and $h_{2}$ vanishes for $\ell = 1$. 
In what follows, we focus on the modes with $\ell \ge 2$ where the odd-parity perturbations are dynamical.
Also, we do not consider the perturbation of the scalar field, because it belongs to the even-parity perturbations. 

In order to eliminate an unphysical degree of freedom, we consider an infinitesimal coordinate transformation: $x^{a} \to x^{a} + \epsilon \xi^{a}$. 
A general infinitesimal transformation for the odd-parity modes can be written as
\begin{align}
    \xi^{a} = \sum_{\ell,m} \Xi_{\ell m}(t,r) E^{ab} \bar{\nabla}_{b} Y_{\ell m}(\theta, \varphi). 
\end{align}
Then, the gauge transformation law for the perturbation variables is given by
\begin{align}
    h_{0} \to h_{0} - \dot \Xi, \qquad
    h_{1} \to h_{1} - \Xi^{\prime} + \frac{2}{r}\Xi, \qquad
    h_{2} \to h_{2} - 2 \Xi,
\end{align}
where a dot and a prime denote the derivatives with respect to $t$ and $r$, respectively. 
For $\ell \ge 2$, we set $h_{2}=0$ to fix the gauge freedom, which is a complete gauge fixing and hence we can legitimately impose it at the action level~\cite{Motohashi:2016prk}. 

The quadratic Lagrangian can be written in terms of a master variable~$\chi_{\ell}$ as follows~\cite{Takahashi:2019oxz}:
\begin{align}
    \frac{2\ell +1 }{2\pi} {\cal L}^{(2)} 
    = 
    \frac{\ell(\ell+1)}{2(\ell-1)(\ell+2)}\sqrt{\frac{B}{A}}
    \brc{b_{1}\dot \chi_{\ell}^{2} - b_{2} \chi_{\ell}^{\prime 2} + b_{3} \dot \chi_{\ell} \chi_{\ell}^{\prime} - \brb{\ell(\ell+1)b_{4} + V_{\rm eff}(r) } \chi_{\ell}^{2}},
    \label{eq:2ndactionchi}
\end{align}
where 
\begin{align}
    b_{1} = \frac{r^{2}{\cal F H}^{2}}{A {\cal FG} + B {\cal J}^{2}},\qquad
    b_{2} = \frac{r^{2}AB{\cal GH}^{2}}{A{\cal FG} + B {\cal J}^{2}},\qquad
    b_{3} = \frac{2r^{2}B{\cal H}^{2}{\cal J}}{A {\cal FG} + B {\cal J}^{2}},\qquad
    b_{4} = {\cal H},
\end{align}
and $V_{\rm eff}(r)$ is given by
\begin{align}
    V_{\rm eff}(r) = 
    r^{2} {\cal H} \brb{b_{2}\sqrt{\frac{B}{A}} \bra{\frac{1}{r^{2}{\cal H}}\sqrt{\frac{A}{B}}}^{\prime}\,}^{\prime} - 2 {\cal H}, 
\end{align}
with ${\cal F}$, ${\cal G}$, ${\cal H}$, and ${\cal J}$ defined by 
\begin{align}
    \begin{split}
    &{\cal F} = 
    2 \bra{F_{2} + \frac{q^{2}}{A}A_{1}},\qquad
    {\cal G} = 
    2\brb{F_{2} - \bra{\frac{q^{2}}{A}+X}A_{1}}, \\
    &{\cal H} = 
    2 \bra{F_{2} - X A_{1}}, \qquad
    {\cal J} = 
    -2q \psi^{\prime} A_{1}. 
    \end{split}
\end{align}
The relation between the master variable and the original perturbation variables can be found in~\cite{Takahashi:2019oxz}.
The existence of the cross term~$b_{3}\dot \chi_{\ell} \chi_{\ell}^{\prime}$ is the crucial difference from the case with $q=0$, $\psi^{\prime}=0$, and/or $A_{1}=0$. 
Indeed, we have $b_{3} \propto {\cal J} \propto q \psi^{\prime} A_{1}$, and hence the cross term vanishes if $q \psi^{\prime} A_{1}=0$. 
However, in the present paper, we do not consider the case where $q \psi^{\prime} A_{1}=0$ because in this case, the equation of motion and consequently the evolution of the odd-parity perturbations are completely the same as those in GR.

Let us proceed with the quadratic Lagrangian~\eqref{eq:2ndactionchi}.
We can eliminate the cross term~$b_{3}\dot \chi_{\ell} \chi_{\ell}^{\prime}$ by introducing a new coordinate~$\tilde{t}$ as follows:
\begin{align}
    \tilde{t} = t + \int \frac{b_{3}}{2b_{2}} {\rm d}r.
    \label{eq:tildetformal}
\end{align}
With this new coordinate, the quadratic Lagrangian becomes
\begin{align}
    {\cal L}^{(2)} \propto
    \tilde{{\cal L}}
    = 
    \frac{1}{2}\sqrt{\frac{B}{A}}
    \brc{
    \tilde{b}_{1} (\partial_{\tilde{t}} \chi_{\ell})^{2}
    - 
    b_{2} \chi_{\ell}^{\prime 2} - \brb{\ell(\ell+1)b_{4}+V_{\rm eff}(r)}\chi_{\ell}^{2}
    }, 
    \label{eq:2ndlagttildediag}
\end{align}
where 
\begin{align}
    \tilde{b}_{1} = b_{1} + \frac{b_{3}^{2}}{4b_{2}}.
\end{align}

Next, we obtain the equation of motion for the odd-parity perturbations. 
Varying the quadratic Lagrangian~\eqref{eq:2ndlagttildediag} with respect to the master variable~$\chi_{\ell}$, 
we obtain the equation of motion as
\begin{align}
    - \partial_{\tilde{t}}^{2}\chi_{\ell} 
    + \frac{b_{2}}{\tilde{b}_{1}} \chi_{\ell}^{\prime \prime}
    + \frac{Ab_{2}B^{\prime} + B(2A b_{2}^{\prime} - b_{2}A^{\prime})}{2AB \tilde{b}_{1}} \chi_{\ell}^{\prime}
    -
    \frac{\ell(\ell+1)b_{4}+V_{\rm eff}} {\tilde{b}_{1}}\chi_{\ell}=0.
\end{align}
We introduce a new coordinate~$\tilde{x}$ and a new variable~$\Psi$ to transform the above equation into the form of a two-dimensional wave equation: 
\begin{align}
    \tilde{x} &= \int \sqrt{\frac{\tilde{b}_{1}}{b_{2}}} {\rm d} r, \label{generalized_tortoise}
    \\
    \Psi_{\ell} 
    &= \frac{\chi_{\ell}}{F(\tilde{x})},
    \label{new_master_variable}
\end{align}
where $F(\tilde{x})$ is given by
\begin{align}
    F(\tilde{x}) = \bra{\frac{A}{B\tilde{b}_{1}b_{2}}}^{1/4}.
\end{align}
Note that $\tilde{x}$ is a generalization of the tortoise coordinate.
Consequently, the equation of motion becomes 
\begin{align}
      \brb{\frac{\partial^{2}}{\partial \tilde{x}^{2}}
    - \frac{\partial^{2}}{\partial \tilde{t}^{2}}
    - V_{\ell}(\tilde{x}) }\Psi_{\ell}
    = 0,
    \label{master_eq}
\end{align}
where $V_{\ell}(\tilde{x})$ is the effective potential defined by 
\begin{align}
    V_{\ell}(\tilde{x}) = \frac{\ell(\ell+1)b_{4}+V_{\rm eff}}{\tilde{b}_{1}} + F \frac{{\rm d}^{2}}{{\rm d}\tilde{x}^{2}} \bra{\frac{1}{F}}.
\end{align}
When we fix the background solution, we can compute the effective potential $V_{\ell}(\tilde{x})$ from the above formula, and hence we can investigate the time evolution of the odd-parity perturbations based on the master equation~\eqref{master_eq}.

It should be noted that one can derive a master equation of the same form even if we do not impose the degeneracy conditions~\eqref{eq:degeneracycons}, as clarified in \cite{Tomikawa:2021pca}.
This is as expected because an extra scalar degree of freedom belongs to the even-parity perturbations and hence does not affect the odd-parity sector.
As mentioned earlier in Sec.~\ref{subsec:BG}, so long as $A_2=-A_1$ is satisfied, the class of higher-order scalar-tensor theories described by the action~\eqref{eq:action_subDHOST} allows for the stealth Schwarzschild-dS solution under the condition~\eqref{existence_condition}.
Moreover, even when $A_2\ne -A_1$ (which happens if we take into account the scordatura term~\cite{Motohashi:2019ymr}), the deviation of the background solution from the stealth Schwarzschild-dS profile is typically very weak and can be negligible at astrophysical scales~\cite{Mukohyama:2005rw,DeFelice:2022qaz}.
Therefore, it is not necessary to impose the degeneracy conditions~\eqref{eq:degeneracycons} for the study of perturbations about the stealth Schwarzschild-dS profile.
Having said that, for concreteness, we focus on the stealth Schwarzschild(-dS) solution in the DHOST theories in the subsequent analyses.

\section{Stealth Schwarzschild solutions}
\label{sec:Schwarzschildcase}

\subsection{Effective metric}

In this section, we consider the stealth Schwarzschild profile as the background solution. 
From the diagonalized quadratic Lagrangian~\eqref{eq:2ndlagttildediag}, we can find the effective metric on which the odd-parity perturbations propagate. 
In what follows, we are interested in the propagation of odd-parity perturbations in the radial direction, and hence we focus on the first two terms in \eqref{eq:2ndlagttildediag} and define a two-dimensional effective metric~$Z_{IJ}$ ($I,J = \{\tilde{t}, r\}$) as
\begin{align}
    \tilde{{\cal L}}_{\rm kin} 
    = \sqrt{\frac{B}{A}} \brb{
    \frac{\tilde{b}_{1}}{2} (\partial_{\tilde{t}} \chi_{\ell})^{2} 
    - 
    \frac{b_{2}}{2} \chi_{\ell}^{\prime 2}
    }
    \eqqcolon
    - \frac{1}{2}
    Z^{IJ} \partial_{I}\chi_{\ell} \partial_{J} \chi_{\ell},
\end{align}
where $Z^{IJ}$ is the inverse of $Z_{IJ}$. 
The component of the effective metric is given by 
\begin{align}
    Z_{IJ} {\rm d}x^{I} {\rm d}x^{J}
    = 
    \sqrt{\frac{A}{B}} 
    \brb{
    - \frac{1}{\tilde{b}_{1}} {\rm d}\tilde{t}^{2} + \frac{1}{b_{2}} {\rm d}r^{2}
    }.
    \label{eq:effectivemetric}
\end{align}
Note that the effective metric is in general different from the background metric, i.e., $Z_{IJ} {\rm d}x^{I} {\rm d}x^{J}\ne \bar{g}_{IJ} {\rm d}x^{I} {\rm d}x^{J}$.
For the stealth Schwarzschild solutions, $Z_{\tilde{t} \tilde{t}}$ becomes
\begin{align}
    Z_{\tilde{t} \tilde{t}} 
    = 
    -\frac{F_{2}(r-r_{\rm s}) - q^{2}A_{1}r_{\rm s}}{2r^{3}(F_{2}+q^{2}A_{1})^{2}}.
    \label{eq:Zttcomp}
\end{align}
For the spacetime described by the effective metric~$Z_{IJ}$, the vector field~$\partial_{\tilde t}$ is a Killing vector field. 
The Killing horizon is located at the radius where $Z_{\tilde{t} \tilde{t}}$ changes its sign. 
From Eq.~\eqref{eq:Zttcomp}, the radius of the Killing horizon, denoted by $r_{\rm g}$,  can be read off as 
\begin{align}
    r_{\rm g} = \bra{1 + \frac{q^{2}A_{1}}{F_{2}}} r_{\rm s} 
    \eqqcolon (1+\zeta)r_{\rm s}.
\end{align}
Since the conditions for no ghost/gradient instabilities are given by~\cite{Takahashi:2021bml}
\begin{align}
    F_{2}>0,\qquad F_{2} + q^{2} A_{1} > 0, 
\end{align}
the Killing horizon $r_{\rm g}$ is positive. 
Note that these conditions imply $\zeta>-1$. 
Note also that $r_{\rm g} > r_{\rm s}$ for $\zeta>0$, while $r_{\rm g} < r_{\rm s}$ for $\zeta<0$. 
The two radii coincide with each other for $q^{2}A_{1}=0$, or equivalently $\zeta = 0$.

\subsection{Characters of a constant-\texorpdfstring{$\tilde{t}$}{tilde t} surface}
\label{subsec:tildetsch}

\begin{figure}
    \centering
    \includegraphics[width=\textwidth]{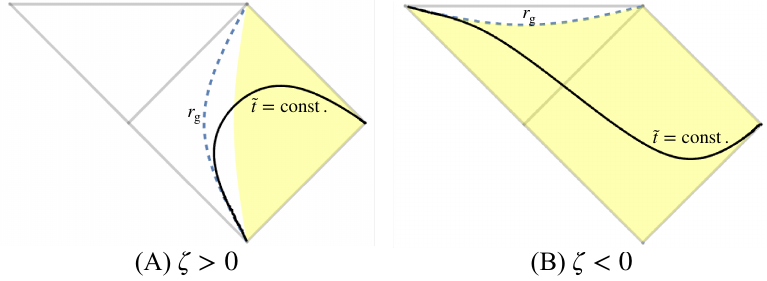}
    \caption{Typical behavior of constant-$\tilde{t}$ surface for (A) $\zeta > 0$ and (B) $\zeta < 0$ embedded in the Penrose diagram of the Schwarzschild spacetime. 
    The black curves represent constant-$\tilde{t}$ surfaces. 
    The constant-$\tilde{t}$ surface is spacelike in the yellow shaded region. 
    }
    \label{fig:consttildetSch}
\end{figure}

Next, we discuss characters of the new time coordinate $\tilde{t}$. 
For the stealth Schwarzschild solutions, 
$\tilde{t}$ can be analytically obtained from Eq.~\eqref{eq:tildetformal} as follows:
\begin{align}
    \tilde{t} 
    = t 
    + 2 \sqrt{\frac{r}{r_{\rm s}}}(r_{\rm s} - r_{\rm g})
    - 
    \frac{1}{\sqrt{r_{\rm s}}}
    \bra{
    r_{\rm g}^{3/2} \log 
    \norm{\frac{\sqrt{r} - \sqrt{r_{\rm g}}}{\sqrt{r} + \sqrt{r_{\rm g}}}}
    - 
    r_{\rm s}^{3/2}
    \log 
    \norm{\frac{\sqrt{r} - \sqrt{r_{\rm s}}}{\sqrt{r} + \sqrt{r_{\rm s}}}}
    } + \tilde{t}_{\rm c}, 
    \label{eq:tildetsch}
\end{align}
where $\tilde{t}_{\rm c}$ is an integration constant. 
Let us investigate whether a constant-$\tilde{t}$ surface is spacelike with respect to the background metric or not. 
To this end, we consider a vector field~$\partial_{\mu}\tilde{t}$ which is normal to a constant-$\tilde{t}$ surface. 
The norm of $\partial_{\mu}\tilde{t}$ associated with the background metric is given by
\begin{align}
    \bar{g}^{\mu \nu} \partial_{\mu}\tilde{t} \partial_{\nu}\tilde{t} = \frac{r(r_{\rm g}^{2} - r r_{\rm s})}{r_{\rm s}(r - r_{\rm g})^{2}}.
\end{align}
Therefore, the constant-$\tilde{t}$ surface is spacelike for $r > r_{\rm g}^{2}/r_{\rm s}$, while it is timelike for $r < r_{\rm g}^{2}/r_{\rm s}$. 
Now, we discuss the relation between the location of the Killing horizon for the odd-parity perturbations~$r_{\rm g}$ and the characteristic radius~$r_{\rm g}^{2}/r_{s}$. 
The Killing horizon~$r_{\rm g}$ is greater than the characteristic radius $r_{\rm g}^{2}/r_{\rm s}$ 
if $r_{\rm s} > r_{\rm g}$, or equivalently $\zeta < 0$. 
Consequently, if we focus on the spacetime in the range~$r>r_{\rm g}$, the constant-$\tilde{t}$ surface is always spacelike.
On the other hand, the Killing horizon~$r_{\rm g}$ is smaller than the characteristic radius~$r_{\rm g}^{2}/r_{\rm s}$
if $r_{\rm s} < r_{\rm g}$, or equivalently $\zeta>0$. 
Therefore, the constant-$\tilde{t}$ surface becomes spacelike in the range~$r>r_{\rm g}^{2}/r_{\rm s}$, while it becomes timelike in the range~$r_{\rm g}< r < r_{\rm g}^{2}/r_{\rm s}$. 
Figure~\ref{fig:consttildetSch} shows the typical behavior of the constant-$\tilde{t}$ surface embedded in the Penrose diagram of the Schwarzschild spacetime. 
The black solid curves are the constant-$\tilde{t}$ surfaces. 
In the yellow shaded region, the constant-$\tilde{t}$ surfaces are spacelike. 

\subsection{Characteristic curves}

In the high-frequency regime, the odd-parity perturbations propagate along the characteristic curves on which either $\tilde{v}=\tilde{t}+\tilde{x}={\rm const}.$ or $\tilde{u}=\tilde{t}-\tilde{x}={\rm const}.$ is satisfied. 
To understand properties of the characteristic curves, we perform a similar analysis as the one in the previous subsection. 
That is, we study the vector fields~$\partial_{\mu} \tilde{u}$ and $\partial_{\mu} \tilde{v}$ which are normal to the characteristic curves. 
The norms of these vector fields with respect to the background metric are given by
\begin{align}
    \bar{g}^{\mu \nu} \partial_{\mu }\tilde{u} \partial_{\nu }\tilde{u} = \frac{r(r_{\rm g} - r_{\rm s})}{r_{\rm s}(\sqrt{r}-\sqrt{r_{\rm g}})^{2}},
    \qquad
    \bar{g}^{\mu \nu} \partial_{\mu }\tilde{v} \partial_{\nu }\tilde{v} = \frac{r(r_{\rm g} - r_{\rm s})}{r_{\rm s}(\sqrt{r}+\sqrt{r_{\rm g}})^{2}},
\end{align}
respectively. 
Therefore, for $r_{\rm g}>r_{\rm s}$ or equivalently $\zeta > 0$, the characteristic curves are timelike, while for $r_{\rm g}<r_{\rm s}$ or equivalently $\zeta < 0$, the characteristic curves are spacelike, i.e., the odd-parity perturbations become superluminal. 
For $\zeta < 0$ case, due to the superluminal propagation, perturbations can propagate from the region in $r_{\rm g} < r < r_{\rm s}$ to that in $r > r_{\rm s}$ (see Appendix~\ref{app:negativezeta}).
\begin{figure}
    \centering
    \includegraphics[width=\textwidth]{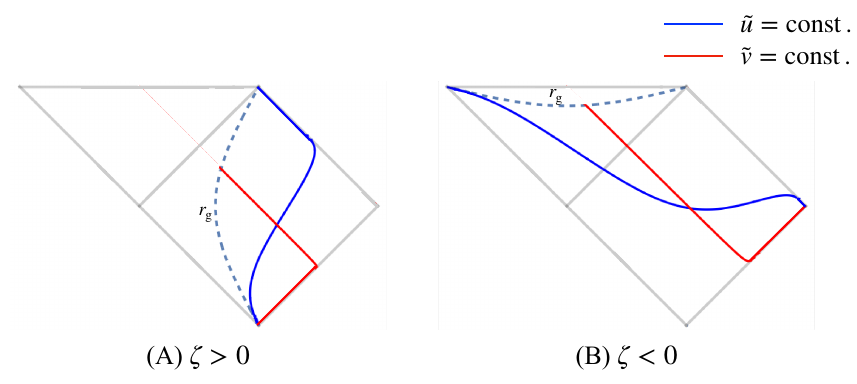}
    \caption{The characteristic curves for (A) $\zeta > 0$ and (B) $\zeta < 0$ embedded in the Penrose diagram of the Schwarzschild spacetime.  
    The red curves and the blues curves represent constant-$\tilde{v}$ curves and constant-$\tilde{u}$ curves, respectively. 
    For (A) $\zeta>0$, the characteristic curves are always timelike, while for (B) $\zeta<0$, the characteristic curves are spacelike, i.e., the odd-parity perturbations are superluminal. 
    }
    \label{fig:uvSchcase}
\end{figure}
Figure~\ref{fig:uvSchcase} shows the characteristic curves embedded in the Penrose diagram of the Schwarzschild spacetime.

\subsection{Equation of motion and QNM frequencies}

Let us study the master equation~\eqref{master_eq} for the case of stealth Schwarzschild solutions.
The generalized tortoise coordinate~$\tilde{x}$ and the new master variable~$\Psi_\ell$ defined in Eqs.~\eqref{generalized_tortoise} and \eqref{new_master_variable} take the form of
\begin{align}
    \tilde{x} &= 
    \sqrt{1+\zeta} 
    \brb{r + r_{\rm g} \log \left|\frac{r}{r_{\rm g}}-1\right|},
    \\
    \Psi_{\ell} &= r \sqrt{2 F_{2}}\bra{\frac{r_{\rm g}}{r_{\rm s}}}^{3/4} \chi_{\ell}.
\end{align}
We note that $\tilde{x} \to -\infty$ as $r \to r_{\rm g}$ and $\tilde{x} \to \infty$ as $r \to \infty$. 
Here, we have chosen the integration constant for $\tilde{x}$ so that $\tilde{x}=0$ at $r=0$. 
The master equation~\eqref{master_eq} is now written as
\begin{align}
    \brb{\frac{\partial^{2}}{\partial \tilde{x}^{2}} - \frac{\partial^{2}}{\partial \tilde{t}^{2}} - V_{\ell}(\tilde{x}) }\Psi_{\ell}
    =0,
    \label{eq:waveequation}
\end{align}
where 
\begin{align}
    V_{\ell}(\tilde{x}) 
    = 
    \frac{1}{1+\zeta}\bra{1-\frac{r_{\rm g}}{r}}
    \brb{
    \frac{\ell(\ell + 1)}{r^{2}}
    - 
    \frac{3r_{\rm g}}{r^{3}}
    }. 
    \label{eq:Veff}
\end{align}
Note that if $\zeta=0$, the above equation reduces to the standard Regge-Wheeler equation in GR.

It should be noted that the master equation~\eqref{eq:waveequation} for the odd-parity perturbations about stealth solutions in the DHOST theory is the same as the one in GR except that the effective potential is multiplied by the factor of $(1+\zeta)^{-1}$ [see Eq.~\eqref{eq:Veff}]. 
Indeed, if we introduce rescaled coordinates~$\tilde{T}$ and $\tilde{X}$ as 
\begin{align}
    \tilde{T} &= \frac{\tilde{t}}{\sqrt{1+\zeta}},
    \label{eq:reslawSch}
    \\
    \tilde{X} &= \frac{\tilde{x}}{\sqrt{1+\zeta}} = r + r_{\rm g} \log \left|\frac{r}{r_{\rm g}}-1\right|,
    \label{eq:Xtilde}
\end{align}
then the master equation~\eqref{eq:waveequation} can be rewritten as 
\begin{align}
    \brb{\frac{\partial^{2}}{\partial \tilde{X}^{2}} - \frac{\partial^{2}}{\partial \tilde{T}^{2}} - \tilde{V}_{\ell}(\tilde{X})}\Psi_{\ell}
    = 0,
    \label{eq:RWequationinrg}
\end{align}
with 
\begin{align}
    \tilde{V}_{\ell}(\tilde{X}) = \bra{1-\frac{r_{\rm g}}{r}}
    \brb{
    \frac{\ell(\ell + 1)}{r^{2}}
    - 
    \frac{3r_{\rm g}}{r^{3}}
    }. 
\end{align} 
Equation~\eqref{eq:RWequationinrg} is nothing but the standard Regge-Wheeler equation in GR if we identify $r_{\rm g}$ as the Schwarzschild radius. 
This implies that we can map a solution for the wave equation in GR to a solution in the DHOST theory: The latter is obtained by just rescaling the coordinates in the former. 
This fact can be used to discuss the QNM frequencies and the power-law tail in the DHOST theory.

Let us first discuss the QNM frequencies. 
Substituting the ansatz~$\Psi_{\ell} = \psi_{\ell}(\tilde{X}) e^{-i \tilde{W} \tilde{T}}$ into Eq.~\eqref{eq:RWequationinrg}, we have
\begin{align}
    \brb{- \frac{{\rm d}^{2}  }{{\rm d}\tilde{X}^{2}} + \tilde{V}_{\ell}(\tilde{X})} \psi_{\ell}(\tilde{X})
    = 
    \tilde{W}^{2} \psi_{\ell}(\tilde{X}).
    \label{eq:schrodingerres}
\end{align}
The QNMs are defined as the modes that are purely ingoing ($\psi_{\ell} \sim e^{-i\tilde{W}\tilde{X}}$) as $r \to r_{\rm g}$, and purely outgoing ($\psi_{\ell} \sim e^{i\tilde{W} \tilde{X}}$) as $r \to \infty$. 
Let $\omega^{\text{Sch}}_{\ell,n}(r_{\rm s})$ be the QNM frequencies for the Schwarzschild spacetime in GR obtained by solving the standard Regge-Wheeler equation, where $n$ is the overtone number. 
For instance, $\omega_{2,0}^{\text{Sch}} = (0.74734 - 0.17792\, i )/r_{\rm s}$ for the $\ell=2$ fundamental mode. 
Also, let $\tilde{W}_{\ell,n}(r_{\rm g})$ be the QNM frequencies obtained by solving Eq.~\eqref{eq:schrodingerres}. 
The relation between $\omega^{\text {Sch}}_{\ell,n}$ and $\tilde{W}_{\ell, n}$ is given by $r_{\rm g}\,\tilde{W}_{\ell, n} = r_{\rm s}\,\omega^{\text{Sch}}_{\ell, n}$. 
From Eq.~\eqref{eq:reslawSch}, we can rewrite the ansatz for $\Psi_{\ell}$ as $\Psi_{\ell} = \psi_{\ell}(\tilde{x}) e^{- \tilde{W} \tilde{t}/\sqrt{1+\zeta}} \eqqcolon \psi_{\ell}(\tilde{x}) e^{-i \omega^{\rm DHOST} \tilde{t}}$. 
Then, the QNM frequencies $\omega^{\rm DHOST}_{\ell,n}$ can be expressed in terms of $\omega^{\text{Sch}}_{\ell,n}$ as 
\begin{align}
    \omega^{\rm DHOST}_{\ell,n} = \frac{r_{\rm s}\, \omega^{\text{Sch}}_{\ell,n}}{r_{\rm s} (1+\zeta)^{3/2}},
    \label{eq:frescaleSch}
\end{align}
where we have used $r_{\rm g} = (1+\zeta)r_{\rm s}$. 
Note that the numerator is the QNM frequencies of the Schwarzschild spacetime in GR in unit of $r_{\rm s}^{-1}$, which we already know.
For instance, for the $\ell=2$ fundamental mode, we have $r_{\rm s}\,\omega_{2,0}^{\text{Sch}} = 0.74734 - 0.17792\,i$, and hence
\begin{align}
    \omega^{\rm DHOST}_{2,0} = \frac{0.74734 - 0.17792\,i}{r_{\rm s} (1+\zeta)^{3/2}}.
\end{align}
Equation~\eqref{eq:frescaleSch} shows that, even if we know QNM frequencies for multiple pairs of $(\ell,n)$ from observations, we can determine only the combination~$r_{\rm s} (1+\zeta)^{3/2}$. 
In this sense, we conclude that there is a degeneracy between $r_{\rm s}$ and $\zeta$. 
Another important consequence is that we can find the QNM frequencies of the stealth Schwarzschild solutions in the DHOST theory from those in GR by applying the formula~\eqref{eq:frescaleSch}.
This is consistent with the result of~\cite{Mukohyama:2023xyf} where the QNM frequencies have been studied based on the effective field theory with a timelike scalar profile~\cite{Mukohyama:2022enj,Mukohyama:2022skk} applied to a static and spherically symmetric black hole background.

Let us now briefly discuss the behavior of the power-law tail in the DHOST theory, assuming that it exists.
It is well known that the power-law tail dominates the waveform of the black hole perturbations after the damped oscillation phase in GR~\cite{Price:1971fb}. 
Now, suppose that the solution to the wave equation~\eqref{eq:RWequationinrg} (i.e., the one rewritten in the form of the standard Regge-Wheeler equation in GR) asymptotically behaves as $\Psi_{\ell} \sim \tilde{T}^{\mathtt{k}}$ at late time, where $\mathtt{k}$ is a negative constant.
Then, by use of Eq.~\eqref{eq:reslawSch}, 
we find that $\Psi_\ell\sim (1+\zeta)^{-\mathtt{k}/2}\tilde{t}^\mathtt{k}\propto \tilde{t}^\mathtt{k}$.
Therefore, if the power-law tail exists in the DHOST theory, we expect that its power would be the same as the one in GR.

Before concluding this subsection, we mention the need for a time-domain analysis. 
In GR, when we obtain the time evolution of the black hole perturbations as a solution of Cauchy problem, the late-time behavior of the black hole perturbations is dominated by a superposition of the QNMs (and the power-law tail). 
On the other hand, in the DHOST (or any other modified gravity) theories, it is nontrivial whether or not the same thing happens because the effective metric for perturbations does not coincide with the background metric in general.
Although we neglect matter fields in the present paper, they exist in reality and their dynamics is determined by the background metric, provided that they are minimally coupled to gravity.
Therefore, 
in order to obtain the time evolution of the perturbations, we should impose initial conditions on a hypersurface which is spacelike with respect to both the background metric and the effective metric. 
In GR, for example, the initial surface is often chosen to be a hypersurface with constant Killing time. 
In the present case of DHOST theories, a portion of a constant-$\tilde{t}$ hypersurface can be timelike with respect to the background metric for $\zeta > 0$. 
When we impose the initial conditions in the region where the constant-$\tilde{t}$ hypersurface is spacelike, the late-time behavior of the perturbations would be dominated by the QMNs with frequencies~$\omega^{\rm{DHOST}}_{\ell,n}$. 
However, when we impose the initial conditions in the region where the constant-$\tilde{t}$ hypersurface is timelike, it is not obvious whether the QNMs dominate the late-time behavior of the perturbations because $\tilde{t}$ cannot be regarded as a physical time coordinate in this case. 
In the next subsection, we show that we can prepare a hypersurface which is spacelike with respect to both the background metric and the effective metric in the region where the constant-$\tilde{t}$ hypersurface is timelike by tilting the constant-$\tilde{t}$ hypersurface in an appropriate manner.

\subsection{Initial value problem and excitations of QNMs}
\label{subsec:initialvproblem}

As we mentioned in Sec.~\ref{subsec:tildetsch}, for $\zeta > 0$, a constant-$\tilde{t}$ surface is timelike with respect to the background metric in the region~$r_{\rm g}< r < r_{\rm g}^{2}/r_{\rm s}$. 
Therefore, in order to discuss the time evolution of the perturbations based on the mater equation~\eqref{eq:waveequation} in a physically sensible manner, we need to choose another initial hypersurface that is spacelike with respect to both the background metric and the effective metric.
Such a formulation of initial value problem has been proposed in \cite{Nakashi:2022wdg}, which we adopt in the following.
In what follows, we focus on the case with $\zeta>0$ and study the $\ell=2$ mode for concreteness (and hence the subscript~$\ell$ will be omitted). 
We analyze the initial value problem for $\zeta < 0$ in Appendix~\ref{app:negativezeta}.

Let us briefly review how we construct an initial surface in the physically sensible formulation proposed in \cite{Nakashi:2022wdg}. 
We introduce new coordinates so that $\tilde{\mathcal{U}} = a\tilde{u}$ and $\tilde{\mathcal{V}} = b \tilde{v} $, where $a$ and $b$ are positive constants. 
\begin{figure}
    \centering
    \includegraphics[width=\textwidth]{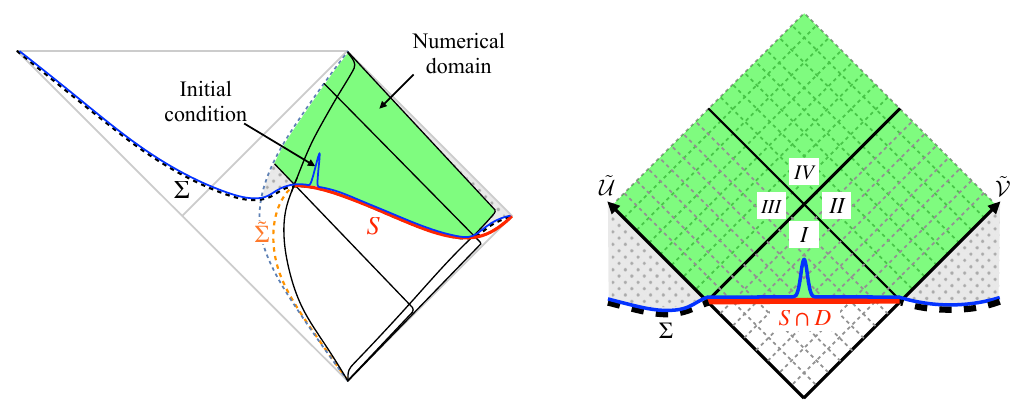}
    \caption{
    Schematic picture of the initial surface~$\Sigma$ (black dashed curve) and the surface~$\tilde{\Sigma}$ (orange dashed curve) which is constructed by tilting the constant-$\tilde{t}$ surface. 
    We put an initial Gaussian wave packet (blue solid curve) on the initial surface~$\Sigma$. 
    We require that~\eqref{req1} the initial surface~$\Sigma$ coincides with the surface~$\tilde{\Sigma}$ in the region~$S$ (red solid curve) within the numerical domain~$D$ (green shaded region). 
    We also require that~\eqref{req2} the initial conditions have a compact support in the region~$S\cap D$, and hence the field vanishes outside the numerical domain (gray shaded region). 
    We further assume that the derivative of the field in the direction perpendicular to $\Sigma$ is zero on the initial surface. 
    }
    \label{fig:inisurface}
\end{figure}
Figure~\ref{fig:inisurface} shows a schematic picture of the initial surface and the numerical domain. 
The left panel shows our numerical setup in the Penrose diagram of the Schwarzschild spacetime, while the right panel shows it in a diagram in which the characteristic curves of the odd-parity perturbations are depicted by 45- and 135-degree straight lines. 
By adjusting the constants~$a$ and $b$, we can make a hypersurface of constant $\tilde{\mathcal{U}} + \tilde{\mathcal{V}}$ be spacelike in the region~$r>r_{\rm B}$ for some $r_{\rm B} < r_{\rm g}^{2}/r_{\rm s}$.
We call the constant-$(\mathcal{\tilde{U}}+\mathcal{\tilde{V}})$ surface $\tilde{\Sigma}$. 
Let $S$ be the region where the hypersurface~$\tilde{\Sigma}$ is spacelike and let $\Sigma$ denote a spacelike hypersurface on which we impose initial conditions and the numerical domain~$D$. 
We impose the following requirements on the hypersurface~$\Sigma$ and the initial conditions:
\begin{enumerate}[(a)]
    \item \label{req1} The initial surface~$\Sigma$ coincides with $\tilde{\Sigma}$ in the region~$S\cap D$.
    \item \label{req2} The initial conditions have a compact support in the region~$S\cap D$.
\end{enumerate}
Since the numerical domain is a part of the causal future of the region~$S$ determined by the characteristic curves of the odd-parity perturbations, in the numerical domain, imposing initial conditions on $\Sigma$ corresponds to imposing initial conditions on $\tilde{\Sigma}$ under the requirement~\eqref{req1}. 
Also, under the requirement~\eqref{req1}, we can regard $\tilde{\mathcal{U}}+\tilde{\mathcal{V}}$ as a physical time in the numerical domain. 
The requirement~\eqref{req2} allows us to obtain the time evolution as follows.
First, we can obtain the time evolution in the region~I in the right panel of Fig.~\ref{fig:inisurface} from the initial data given in the region~$S\cap D$.
Then, when we study the time evolution in the regions~II and III, we can use the requirement~\eqref{req2} to set $\Psi=0$ on both the right boundary of region~II and the left boundary of region~III.
Once we obtain the solution in the regions~II and III, it is straightforward to compute the time evolution in the region~IV.
Thus, we can obtain the time evolution of the field in the whole numerical domain.

\begin{figure}
    \centering
    \includegraphics[width=0.6\textwidth]{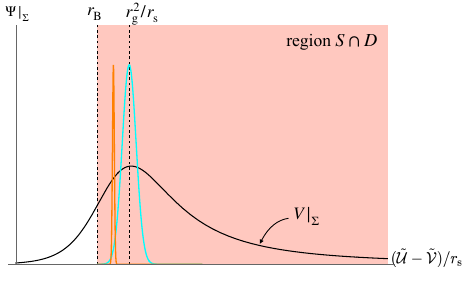}
    \caption{
    Schematic picture of the initial field profiles. 
    The cyan and the orange curves are the Gaussian wave packets with $\sigma = r_{\rm s}$ and $\sigma = 0.1 r_{\rm s}$, respectively. 
    The horizontal axis is the value of $(\tilde{\mathcal{U}}-\tilde{\mathcal{V}})/r_{\rm s}$ on the initial surface~$\Sigma$.
    In practice, we truncate the Gaussian function to have a compact support within the region~$S \cap D$ and choose 
    the center of the wave packet so that its support overlaps with the region~$r_{\rm B} < r < r_{\rm g}^{2}/r_{\rm s}$. 
    The black curve is the schematic plot of the effective potential~$V(\tilde{\mathcal{U}}, \tilde{\mathcal{V}})$
    on the initial surface, which is meant to show the position of the potential peak (and hence its height does not have any particular meaning).
    }
    \label{fig:initial_field_profile}
\end{figure}

We consider a Gaussian wave packet as the initial field profile:
\begin{align}
    \Psi|_{\Sigma}
    =\Psi(C_0-\tilde{\mathcal{V}},\tilde{\mathcal{V}})|_{\tilde{\Sigma}}
    =e^{-\frac{1}{2} \bra{ \frac{\tilde{\mathcal{V}}-\tilde{\mathcal{V}}_{0}}{\sigma} }^{2}},
    \label{eq:inicon}
\end{align}
where $\sigma$ and $\tilde{\mathcal{V}}_{0}$ are the width of the Gaussian wave packet and its peak position, respectively.
It should be noted that we truncate the Gaussian profile in a finite region in our actual computations so that the initial data have a compact support within the region~$S\cap D$. 
We recall that $\tilde{\mathcal{U}}+\tilde{\mathcal{V}}$ takes a constant value (which we denote by $C_0$) on the initial surface within the numerical domain thanks to the requirement~\eqref{req1}, and hence it makes sense to define the initial data as in Eq.~\eqref{eq:inicon}.
We choose $\sigma$ and $\tilde{\mathcal{V}}_0$ so that the support of the initial field profile overlaps with the region~$r_{\rm B}<r<r_{\rm g}^{2}/r_{\rm s}$, where the surface of constant $\tilde{\mathcal{U}}+\tilde{\mathcal{V}}$ is spacelike and the constant-$\tilde{t}$ surface is timelike (see Fig.~\ref{fig:initial_field_profile}).
Also, regarding the initial condition for the derivative, we impose $(\partial_{\tilde{\mathcal{U}}} + \partial_{\tilde{\mathcal{V}}}) \Psi|_{\Sigma} = 0$.

Let us now explain how we solve the master equation~\eqref{eq:waveequation} under the initial conditions mentioned above.
Expressing the master equation~\eqref{eq:waveequation} in terms of $\tilde{\mathcal{U}}$ and $\tilde{\mathcal{V}}$, we have 
\begin{align}
    -4 \frac{\partial^{2} \Psi}{\partial \tilde{\mathcal{U}} \partial\tilde{\mathcal{V}}} = \frac{V(\tilde{\mathcal{U}}, \tilde{\mathcal{V}})}{ab} \Psi.
    \label{eq:waveintutv}
\end{align}
We discretize the coordinates $\tilde{\mathcal{U}}$ and $\tilde{\mathcal{V}}$ as $\{\tilde{\mathcal{U}}_{i},\tilde{\mathcal{V}}_{j}\}$ where $i,j=0,1,2,\cdots$.
Note that the grid width~$h$ is assumed to be uniform: $h = \tilde{\mathcal{U}}_{i+1}-\tilde{\mathcal{U}}_{i} = \tilde{\mathcal{V}}_{j+1}-\tilde{\mathcal{V}}_{j}$.
Also, we introduce shorthand notations~$\Psi_{i,j}=\Psi(\tilde{\mathcal{U}}_{i},\tilde{\mathcal{V}}_{j})$ and $V_{i,j}=V(\tilde{\mathcal{U}}_{i},\tilde{\mathcal{V}}_{j})$. 
Then, we apply the discretization scheme introduced by~\cite{Gundlach:1993tp}, and the master equation~\eqref{eq:waveintutv} is simply discretized as 
\begin{align}
    \Psi_{i+1,j+1} = \Psi_{i+1,j} + \Psi_{i,j+1} - \Psi_{i,j} - \frac{h^{2}}{8}\frac{V_{i,j}}{ab}\brb{\Psi_{i+1,j} + \Psi_{i,j+1}} + \mathcal{O}(h^{4}).
    \label{eq:disceq}
\end{align}
The initial field profile~\eqref{eq:inicon} can be implemented as
\begin{align}
    \Psi_{i,j}=e^{-\frac{1}{2} \bra{ \frac{\tilde{\mathcal{V}}_j-\tilde{\mathcal{V}}_{0}}{\sigma} }^{2}}
    \Theta(\tilde{\mathcal{V}}-\tilde{\mathcal{V}}_{j_1})\,\Theta(\tilde{\mathcal{V}}_{j_2}-\tilde{\mathcal{V}}), \qquad
    (\tilde{\mathcal{U}}_i,\tilde{\mathcal{V}}_j)\in S\cap D.
    \label{eq:inicon_disc}
\end{align}
Here, to make the truncation explicit, we have inserted the step functions (denoted by $\Theta$) so that $\Psi$ is nonvanishing only for $\tilde{\mathcal{V}}_{j_1}\le \tilde{\mathcal{V}}\le \tilde{\mathcal{V}}_{j_2}$ on the initial surface.
Also, the condition~$(\partial_{\tilde{\mathcal{U}}} + \partial_{\tilde{\mathcal{V}}}) \Psi|_{\Sigma} = 0$ yields
\begin{align}
    \Psi_{i,j} = \Psi_{i+1,j+1} + \mathcal{O}(h^{4}), \qquad
    (\tilde{\mathcal{U}}_i,\tilde{\mathcal{V}}_j)\in S\cap D,
\end{align}
and hence we have
\begin{align}
    \Psi_{i+1,j+1} = \frac{1}{2} \brb{\Psi_{i+1,j} + \Psi_{i,j+1}} - \frac{h^{2}}{16} \frac{V_{i,j}}{ab} \brb{\Psi_{i+1,j} + \Psi_{i,j+1}} + \mathcal{O}(h^{4}), \qquad
    (\tilde{\mathcal{U}}_i,\tilde{\mathcal{V}}_j)\in S\cap D.
    \label{eq:discder}
\end{align}
Combining the discretized equation~\eqref{eq:disceq} as well as the initial conditions~\eqref{eq:inicon_disc} and \eqref{eq:discder}, we can obtain the solution for $\Psi$ in the whole numerical domain.

A caveat should be added here.
One may think that a constant-$\phi$ surface would be a good candidate for the initial surface as it is spacelike with respect to both the background metric and the effective metric.
However, if we choose a constant-$\phi$ surface as the initial surface, it is nontrivial how to implement initial conditions in our numerical scheme which is based on a double-null grid, since a constant-$\phi$ surface cannot be described by a linear function of the null coordinates.
This is the reason why we have chosen the initial surface~$\Sigma$ as above.\footnote{If one would like to choose a constant-$\phi$ surface as the initial surface, then one needs a numerical scheme that is more suitable for solving the differential equation based on the constant-$\phi$ foliation.}

\begin{figure}
    \centering
    \includegraphics[width=0.7\textwidth]{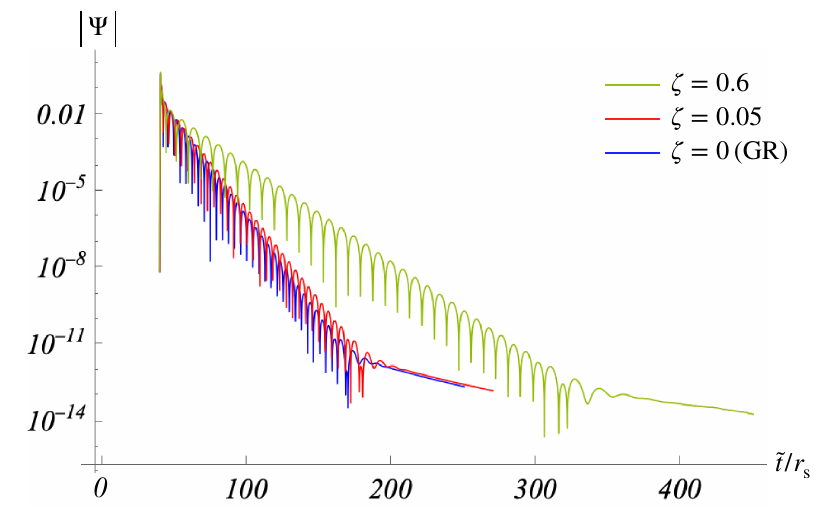}
    \caption{The time evolution of the odd-parity perturbations for $\zeta=0.6$ (green curve, top), $\zeta=0.05$ (red curve, middle), and $\zeta=0$ (blue curve, bottom). 
    The width of the initial Gaussian wave packet is $\sigma = 0.1 r_{\rm s}$. 
    }
    \label{fig:timevoSch}
\end{figure}
Figure~\ref{fig:timevoSch} shows the time evolution of the odd-parity perturbations for $\zeta=0.6$ (green curve, top), $\zeta=0.05$ (red curve, middle), and $\zeta=0$, i.e., GR (blue curve, bottom) for the initial Gaussian wave packet with $\sigma = 0.1 r_{\rm s}$. 
The observer is located at $\tilde{x} = 40 \:r_{\rm s}$. 
The initial Gaussian wave packet first reaches the observer almost unscattered, and then the ringdown phase follows, as can be seen in Fig.~\ref{fig:timevoSch}.

In order to confirm that the frequencies in the ringdown phase are QNM frequencies, we fit the numerical waveform with a superposition of the QNMs. 
We introduce the following fitting model~$\psi_{N}(\tilde{t})$:
\begin{align}
    \psi_{N}(\tilde{t}) = \sum_{n=0}^{N} \alpha_{n} e^{-i \brb{ \mu\, \omega^{\text{Sch}}_{n} (\tilde{t}-\tilde{t}_{\rm peak}) / r_{\rm s} + \beta_{n}}} + c.c., \qquad \tilde{t} \in 
    [\tilde{t}_{0},\tilde{t}_{\rm end}], 
    \label{eq:fittingmodel}
\end{align}
where $n$ labels the overtones and $N$ is the maximum overtone number used in the fitting.
Here, $\alpha_n$ and $\beta_n$ are real parameters corresponding to the amplitude and the phase, respectively, and $\mu$ is a real parameter characterizing the deviation from the QNM frequencies in GR.
Also, $t_{\rm peak}$ denotes the time at which the numerical waveform~$\Psi(\tilde{t})$ takes the maximum value after the initial Gaussian wave packet passes through the observer. 
For the fitting analysis, we use the numerical waveform in the interval~$[\tilde{t}_{0}, \tilde{t}_{\rm end}]$ where $\tilde{t}_{0}$ and $\tilde{t}_{\rm end}$ are free parameters satisfying $\tilde{t}_{\rm peak} \le \tilde{t}_{0} < \tilde{t}_{\rm end}$. 
In our fits, we use the Mathematica function~$\mathtt{NonlinearModelFit}$. 
The amplitude~$\alpha_{n}$ and the phase~$\beta_{n}$ are fitting
parameters, and we find best-fit values of these parameters. 
Note that the parameter~$\mu$ is fixed in the fitting analysis for this section. 
Once we obtain a best-fit function~$\psi_{N}(\tilde{t})$, we evaluate the goodness of the fit by calculating the mismatch~$\mathcal{M}$ defined by
\begin{align}
    \mathcal{M} = 1 - \frac{\langle\Psi | \psi_{N}\rangle}{\sqrt{\langle\Psi | \Psi \rangle \langle\psi_{N} | \psi_{N}\rangle }},
\end{align}
where the scalar product is defined as
\begin{align}
    \langle f | g \rangle = \int_{\tilde{t}_{0}}^{\tilde{t}_{\rm end}} f(\tilde{t}) g^{*}(\tilde{t}) \;{\rm d} \tilde{t},
    \label{eq:innerproduct}
\end{align}
for arbitrary two complex functions~$f$ and $g$, with an asterisk denoting the complex conjugation. 
Note that the fitting model~$\psi_{N}(\tilde{t})$ and the numerical waveform~$\Psi(\tilde{t})$ are real because we impose real initial conditions, and hence the complex conjugate in Eq.~\eqref{eq:innerproduct} is of no particular significance. 
We have kept the complex conjugate just to follow the convention in the literature.
For the value of $\mu$ which is fixed, we consider the following two cases:
\begin{enumerate}
\renewcommand{\theenumi}{\roman{enumi}}
\renewcommand{\labelenumi}{(\theenumi)}
    \item \label{way1} 
    $\mu=r_{\rm s}(1+\zeta)^{-3/2}$:
    We assume the value of $\mu$ as $\mu = r_{\rm s}\,(1+\zeta)^{-3/2}$ with fixed $\zeta$, and find the best-fit parameters~$\alpha_{n}$ and $\beta_{n}$. 
    This case corresponds to the situation where 
    we fit the numerical waveform with a superposition of QNMs with the frequencies~$\omega^{\rm DHOST}_{n}$ in the DHOST theory . 
    \item \label{way2} 
    $\mu=r_{\rm s}$: 
    We assume the value of $\mu$ as $\mu = r_{\rm s}$, and find the best-fit parameters~$\alpha_{n}$ and $\beta_{n}$.  
    This case corresponds to the situation where we fit the numerical waveform with a superposition of QNMs with frequencies~$\omega^{\text{Sch}}_{n}$ in GR.
\end{enumerate}
\begin{figure}
    \centering
    \includegraphics[width=\textwidth]{
    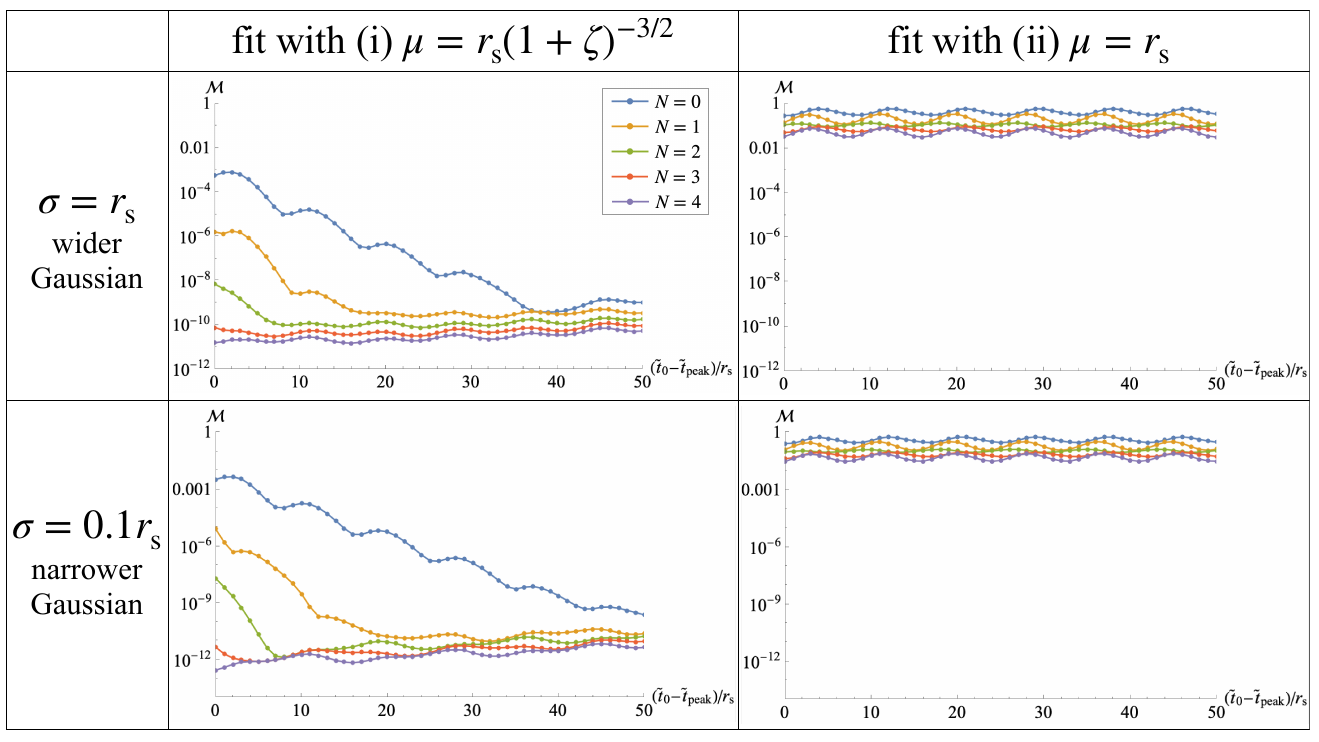
    }
    \caption{The mismatch for $\zeta = 0.6$ between the numerical waveform~$\Psi(\tilde{t})$ and the fitting model~$\psi_{N}(\tilde{t})$ defined by Eq.~\eqref{eq:fittingmodel}. 
    The left panels are calculated with (\ref{way1})~$\mu=r_{\rm s}(1+\zeta)^{-3/2}$, while the right panels are calculated with (\ref{way2})~$\mu=r_{\rm s}$. 
    In addition, the upper panels are the mismatch for $\sigma = r_{\rm s}$, i.e., the wider initial Gaussian wave packet, while the lower panels are the mismatch for $\sigma = 0.1 r_{\rm s}$, i.e., the narrower initial Gaussian wave packet. 
    }
    \label{fig:mismatchSch}
\end{figure}

Figure~\ref{fig:mismatchSch} shows the mismatch for $\zeta = 0.6$. 
The left panels are the mismatch calculated with (\ref{way1})~$\mu = r_{\rm s}(1+\zeta)^{-3/2}$ for $\sigma=r_{\rm s}$ (upper panel) and $\sigma=0.1 r_{\rm s}$ (lower panel), respectively. 
As the number of $N$ increases, the minimum of mismatch decreases. 
When we fit the numerical waveform with only the fundamental mode, i.e., $N=0$ (blue curve), the mismatch gets smaller as $\tilde{t}_{0}$ increases. 
This implies that the waveform near the peak time~$\tilde{t}_{\rm peak}$ is dominated by the overtones. 
Indeed, when we take into account the higher overtones, the value of $\tilde{t}_0$ that minimizes the mismatch gets closer to $\tilde{t}_{\rm peak}$.  
The right panels in Fig.~\ref{fig:mismatchSch} show the mismatch calculated with (\ref{way2})~$\mu = r_{\rm s}$. 
Unlike the DHOST fitting~(\ref{way1}), for all $N$, the mismatch takes almost constant values. 
In particular, we see that the mismatch for $N=0$ does not decrease at late time in the GR fitting~\eqref{way2}.
This reflects the inconsistency between the numerical waveform and the fitting model: We are now fitting the waveform for the DHOST theory by a superposition of the QNMs in GR.
Thus, the right panels in Fig.~\ref{fig:mismatchSch} explicitly show that the the QNMs with the frequencies in GR do not well describe the numerical waveform for the DHOST theory.

\begin{figure}
    \centering  \includegraphics[width=\textwidth]{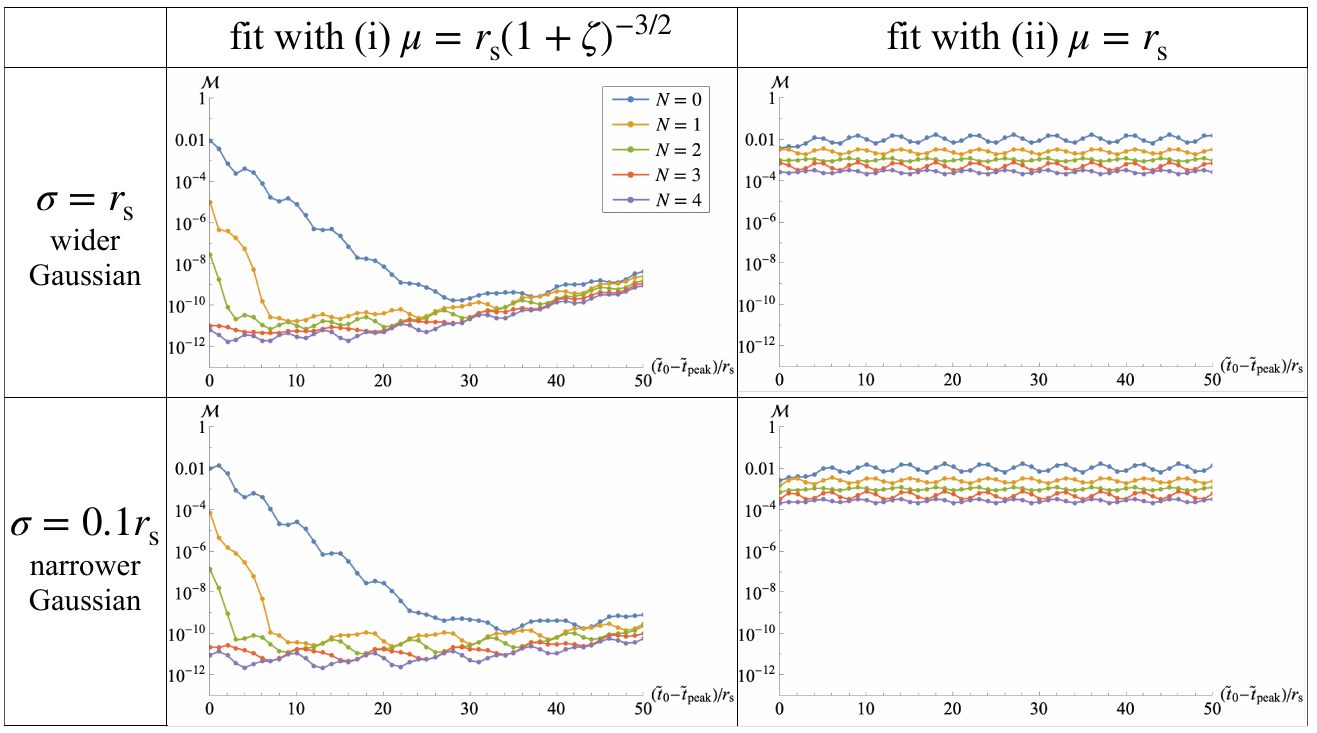}
    \caption{The mismatch for $\zeta = 0.05$ between the numerical waveform~$\Psi(\tilde{t})$ and the fitting model~$\psi_{N}(\tilde{t})$ defined by Eq.~\eqref{eq:fittingmodel}. 
    The values of the parameters in each panel are the same as those in Fig.~\ref{fig:mismatchSch}. 
    }
    \label{fig:mismatch005all}
\end{figure}

Figure~\ref{fig:mismatch005all} shows the mismatch for $\zeta=0.05$. 
As in the case of $\zeta = 0.6$, when we calculate the mismatch with (\ref{way1})~$\mu = r_{\rm s}(1+\zeta)^{-3/2}$, as $N$ increases, the minimum of mismatch decreases and the value of $\tilde{t}_0$ at the minimum gets closer to $\tilde{t}_{\rm peak}$. 
On the other hand, when we calculate the mismatch with (\ref{way2})~$r=r_{\rm s}$, it can be seen that the numerical waveform is not well described with the QNMs in GR. 

From these results, we conclude that the superposition of the QNMs in the DHOST theory~\eqref{eq:frescaleSch} is consistent with the numerical waveform and the QNMs are excited in the physically sensible initial value problem. 
We have also performed the fitting analysis keeping $\mu$ unfixed and found that the best-fit value of $\mu$ is consistent with that of the DHOST theory, i.e., $\mu = r_{\rm s}(1+\zeta)^{-3/2}$.

\section{Stealth Schwarzschild-de Sitter solutions}
\label{sec:SdScase}

In this section, we consider the stealth Schwarzschild-dS profile as the background solution. 
The background metric is given by the Schwarzschild-dS metric: 
\begin{align}
    A(r) = B(r) 
    = - \frac{\Lambda}{3r} \bra{ r^{3} - \frac{3}{\Lambda}r + \frac{3r_{\rm s}}{\Lambda} }
    \eqqcolon 
    - \frac{\Lambda}{3r} \Delta(r),
\end{align}
with $r_{\rm s}$ and $\Lambda$ being positive constants.
Since $\Delta(r)$ is a cubic polynomial in $r$, it can be factorized as $\Delta(r) = (r-r_{-})(r-r_{\rm e})(r-r_{\rm c})$. 
Here, the three roots 
are given by 
\begin{align}
    r_{-}
    &= 
    \frac{2}{\sqrt{\Lambda}} \cos\brb{\frac{1}{3}\cos^{-1}\bra{-\frac{3 r_{\rm s} \sqrt{\Lambda}}{2}}+\frac{2\pi}{3}},
    \\
    r_{\rm e}
    &= 
    \frac{2}{\sqrt{\Lambda}} \cos\brb{\frac{1}{3}\cos^{-1}\bra{-\frac{3 r_{\rm s} \sqrt{\Lambda}}{2}}+\frac{4\pi}{3}},
    \\
    r_{\rm c}
    &= 
    \frac{2}{\sqrt{\Lambda}} \cos\brb{\frac{1}{3}\cos^{-1}\bra{-\frac{3 r_{\rm s} \sqrt{\Lambda}}{2}}},
\end{align}
respectively. 
We note that the three roots are real if 
\begin{align}
  r_{\rm s} \sqrt{\Lambda} < \frac{2}{3},
  \label{eq:conexfornull}
\end{align}
is satisfied, and we have labeled these roots so that $r_{-}<0<r_{\rm e}<r_{\rm c}$.  
Therefore, the event horizon is located at $r=r_{\rm e}$ and the cosmological horizon is located at $r_{\rm c}$. 
For small $r_{\rm s}\sqrt{\Lambda}$, the three roots above are expanded as
    \begin{align}
    r_{-}&=-\sqrt{\frac{3}{\Lambda}}\brb{1+\frac{r_{\rm s}}{2}\sqrt{\frac{\Lambda}{3}}-\frac{r_{\rm s}^2\Lambda}{8}+\frac{r_{\rm s}^3}{6}\sqrt{\frac{\Lambda^3}{3}}+{\cal O}(r_{\rm s}^4\Lambda^2)}, \\
    r_{\rm e}&=r_{\rm s}\brb{1+\frac{1}{3}r_{\rm s}^2\Lambda+{\cal O}(r_{\rm s}^4\Lambda^2)}, \\
    r_{\rm c}&=\sqrt{\frac{3}{\Lambda}}\brb{1-\frac{r_{\rm s}}{2}\sqrt{\frac{\Lambda}{3}}-\frac{r_{\rm s}^2\Lambda}{8}-\frac{r_{\rm s}^3}{6}\sqrt{\frac{\Lambda^3}{3}}+{\cal O}(r_{\rm s}^4\Lambda^2)}.
    \end{align}

\subsection{Effective metric}

For the stealth Schwarzschild-dS solutions, we can define the effective metric in the same manner as in the case of stealth Schwarzschild solutions. 
From the general expression~\eqref{eq:effectivemetric} for the effective metric~$Z_{IJ}$, the $\tilde{t} \tilde{t}$-component can be read off as
\begin{align}
    Z_{\tilde{t}\tilde{t}}
    =
    \frac{1}{2F_{2}(1+\zeta)^{2}r^{2}}
    \brb{1-\frac{(1+\zeta)r_{\rm s}}{r}-\frac{(1+\zeta)\Lambda}{3}r^{2}}.
\end{align}
Introducing a parameter~$\Lambda_{\rm g}=(1+\zeta)\Lambda$, we have
\begin{align}
    Z_{\tilde{t}\tilde{t}}
    =
    -\frac{\Lambda_{\rm g}}{6F_{2}(1+\zeta)^{2}r^{3}}\bra{r^{3}-\frac{3}{\Lambda_{\rm g}}r + \frac{3r_{\rm g}}{\Lambda_{\rm g}}}
    \eqqcolon
    -\frac{\Lambda_{\rm g}}{6F_{2}(1+\zeta)^{2}r^{3}}\Delta_{\rm g}(r).
\end{align}
Therefore, for the odd-parity perturbations, the locations of the Killing horizons are determined by the roots for $\Delta_{\rm g}(r)=0$. 
The roots are given by
\begin{align}
    \tilde{r}_{-}
    &= 
    \frac{2}{\sqrt{\Lambda_{\rm g}}} \cos\brb{\frac{1}{3}\cos^{-1}\bra{-\frac{3 r_{\rm g} \sqrt{\Lambda_{\rm g}}}{2}}+\frac{2\pi}{3}},
    \\
    \tilde{r}_{\rm e}
    &= 
    \frac{2}{\sqrt{\Lambda_{\rm g}}} \cos\brb{\frac{1}{3}\cos^{-1}\bra{-\frac{3 r_{\rm g} \sqrt{\Lambda_{\rm g}}}{2}}+\frac{4\pi}{3}},
    \\
    \tilde{r}_{\rm c}
    &= 
    \frac{2}{\sqrt{\Lambda_{\rm g}}} \cos\brb{\frac{1}{3}\cos^{-1}\bra{-\frac{3 r_{\rm g} \sqrt{\Lambda_{\rm g}}}{2}}}.
\end{align}
We note that these three roots are real if $r_{\rm g}\sqrt{\Lambda_{\rm g}}<2/3$, i.e.,
\begin{align}
    r_{\rm s} \sqrt{\Lambda} < \frac{2}{3(1+\zeta)^{3/2}},
    \label{eq:conexforgraviton}
\end{align}
is satisfied, and we have labeled these roots so that $\tilde{r}_{-} < 0 < \tilde{r}_{\rm e} < \tilde{r}_{\rm c}$. 

In the present paper, we consider the spacetime region satisfying $\Delta_{\rm g}(r)>0$, i.e., the static region for the odd-parity perturbations. 
The static region associated with the background metric is defined by $\Delta(r)>0$, and the relation to the static region for the effective metric depends on the value of $\zeta$ and $\Lambda$.
For $\zeta>0$ and $r_{\rm s}\sqrt{\Lambda} < 2(1+\zeta)^{-3/2}/3$, both the effective metric and the background metric have the static region.
However, for $2 (1+\zeta)^{-3/2}/3 < r_{\rm s}\sqrt{\Lambda} < 2 / 3$, only the background metric has the static region. 
\begin{figure}
    \centering
    \includegraphics[width= 0.7\textwidth]{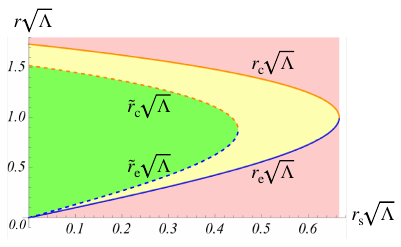}
    \caption{
    Relation between the static region associated with the effective metric and the one associated with the background metric for $\zeta>0$.
    The former (green shaded region) exists for $r_{\rm s}\sqrt{\Lambda} < 2(1+\zeta)^{-3/2}/3$, while the latter (yellow shaded region) exists for $r_{\rm s}\sqrt{\Lambda} < 2/3$.
    }
    \label{fig:staticregion}
\end{figure}
Figure~\ref{fig:staticregion} shows the schematic picture of the relation between the static regions for the effective metric and the background metric in the $\zeta>0$ case. 
The solid blue and orange curves respectively correspond to the event horizon and the cosmological horizon of the background spacetime, while the dashed blue and orange curves correspond to the Killing horizons for the odd-parity perturbations. 
The green shaded region is the static region for 
both the effective and background metrics, while the yellow shaded region is the static region for the background metric only.

\subsection{Characters of a constant-\texorpdfstring{$\tilde{t}$}{tildet t} surface}

We examine the characters of a constant-$\tilde{t}$ surface. 
For the stealth Schwarzschild-dS solutions, the coordinate~$\tilde{t}$ defined by \eqref{eq:tildetformal} reads
\begin{align}
    \tilde{t} = t - \int \frac{\zeta (3r)^{3/2}\sqrt{3r_{\rm s} + \Lambda r^{3}}}{(1+\zeta)\Lambda^{2}\Delta(r) \Delta_{\rm g}(r)}\, {\rm d}r,
\end{align}
though we cannot express it in a simple analytic form unlike the case of the stealth Schwarzschild solutions.
We investigate the global structure of a constant-$\tilde{t}$ surface in the background spacetime. 
To this end, we consider a vector field~$\partial_{\mu} \tilde{t}$ which is normal to the constant-$\tilde{t}$ surface. 
For the background Schwarzschild-dS metric, the norm of $\partial_{\mu} \tilde{t}$ is given by
\begin{align}
    \bar{g}^{\mu \nu} \partial_{\mu} \tilde{t} \,\partial_{\nu}\tilde{t} = 
    \frac{3r}{\Lambda \Delta_{\rm g}(r)} \bra{r^{3} - \frac{3}{(1+\zeta)^{2} \Lambda}r + \frac{3r_{\rm s}}{\Lambda}}.
    \label{eq:normtsds}
\end{align}
Since we focus only on the region in which $\Delta_{\rm g}(r)>0$, the sign of $\bar{g}^{\mu \nu} \partial_{\mu} \tilde{t} \,\partial_{\nu}\tilde{t}$ is determined by the sign of the function in the parentheses in Eq.~\eqref{eq:normtsds}. 
\begin{figure}
    \centering    
    \includegraphics[width=0.9 \textwidth]{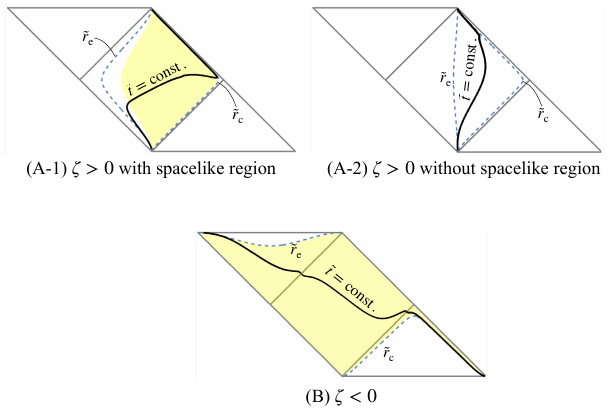}
    \caption{
    Typical plots of constant-$\tilde{t}$ surfaces in the Penrose diagram of the Schwarzschild-dS spacetime. 
    The black curves represent the constant-$\tilde{t}$ surfaces. 
    The constant-$\tilde{t}$ surfaces are spacelike in the yellow shaded region. 
    For (A-1)~$\zeta > 0$, when the parameter $\zeta$ satisfies Eq.~\eqref{eq:sdsspacelikecon}, the constant-$\tilde{t}$ surface can be spacelike in a finite region. 
    For (A-2)~$\zeta > 0$ when the parameter $\zeta$ violates Eq.~\eqref{eq:sdsspacelikecon}, the constant-$\tilde{t}$ surface is always timelike. 
    For (B)~$\zeta < 0$, the constant-$\tilde{t}$ surface is always spacelike. }
    \label{fig:consttildetsds}
\end{figure}
In Fig.~\ref{fig:consttildetsds}, we show typical plots of constant-$\tilde{t}$ surfaces in the Penrose diagram of the Schwarzschild-dS spacetime.\footnote{
In Fig.~\ref{fig:consttildetsds} and Fig.~\ref{fig:uvSdScase}, in the $\zeta<0$ case, a constant-$\tilde{t}$ surface and a characteristic curves apparently become null near both the event horizon and the cosmological horizon, but in fact these are spacelike everywhere. 
This behavior is an artifact caused by the coordinate system we have used. 
For the stealth Schwarzschild-dS solutions, we have defined different double null coordinates in each block of the Penrose diagram and drawn the curves for each block. Then, we have glued the diagrams of each block using the method proposed in~\cite{Walker1970}. 
On the other hand, for the stealth Schwarzschild solutions, we can define a single coordinate system which covers the whole spacetime in a simple analytic form, and hence the coordinate~$\tilde{t}$ and the characteristic curves are written in an analytic form, which has led to the smooth curves in Fig.~\ref{fig:consttildetSch} and Fig.~\ref{fig:uvSchcase}. 
}
The constant-$\tilde{t}$ surfaces are spacelike in the yellow shaded region. 
For $\zeta > 0$, the constant-$\tilde{t}$ surface has a spacelike region within $\tilde{r}_{\rm e}<r<\tilde{r}_{\rm c}$ if 
\begin{align}
    r_{\rm s} \sqrt{\Lambda} < \frac{2}{3(1+\zeta)^{3}},
    \label{eq:sdsspacelikecon}
\end{align}
is satisfied [see (A-1) in Fig.~\ref{fig:consttildetsds}]. 
If the condition~\eqref{eq:sdsspacelikecon} is violated for $\zeta > 0$, the constant-$\tilde{t}$ surface is always timelike [see (A-2) in Fig.~\ref{fig:consttildetsds}]. 
On the other hand, for $\zeta<0$, the constant-$\tilde{t}$ surface is always spacelike [see (B) in Fig.~\ref{fig:consttildetsds}].

\subsection{Characteristic curves}

\begin{figure}
    \centering
    \includegraphics[width=\textwidth]{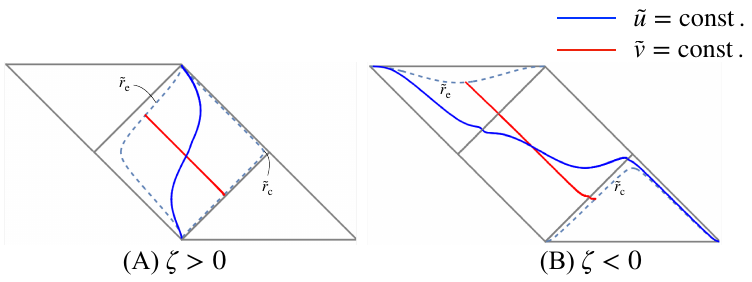}
    \caption{The characteristic curves for (A)~$\zeta > 0$ and (B)~$\zeta < 0$ in the Penrose diagram of the Schwarzschild-dS spacetime.  
    The red curves and the blues curves represent the constant-$\tilde{v}$ curves and the constant-$\tilde{u}$ curves, respectively. }
    \label{fig:uvSdScase}
\end{figure}

As in the case of the stealth Schwarzschild solutions, the odd-parity perturbations propagate along a constant-$\tilde{u}$ curve or a constant-$\tilde{v}$ curve. 
Here, we investigate the vector fields which are normal to the constant-$\tilde{u}$ curve and the constant-$\tilde{v}$ curve: $\partial_{\mu} \tilde{u}$ and $\partial_{\mu} \tilde{v}$. 
For the background Schwarzschild-dS metric, the norm of these vector fields are given by
\begin{align}
    \bar{g}^{\mu \nu} \partial_{\mu} \tilde{u}\, \partial_{\nu} \tilde{u} &= 
    \frac{3\zeta r}{\Lambda_{\rm g}^{2} \Delta_{\rm g}(r)^{2}} \bra{ \sqrt{3 r_{\rm g} + \Lambda_{\rm g}r^{3}} + \sqrt{3 r} }^{2},
    \\
    \bar{g}^{\mu \nu} \partial_{\mu} \tilde{v}\, \partial_{\nu} \tilde{v} &= 
    \frac{3\zeta r}{\Lambda_{\rm g}^{2} \Delta_{\rm g}(r)^{2}} \bra{ \sqrt{3 r_{\rm g} + \Lambda_{\rm g}r^{3}} - \sqrt{3 r} }^{2}.
\end{align}
The sign of each norm is determined by the sign of the parameter $\zeta$. 
The characteristic curves are timelike for $\zeta>0$, while they are spacelike for $\zeta<0$. 
That is, for $\zeta < 0$, the odd-parity perturbations become superluminal. 
Figure~\ref{fig:uvSdScase} shows the characteristic curves of the odd-parity perturbations in the Penrose diagram of the Schwarzschild-dS spacetime.

\subsection{Equation of motion and QNM frequencies}

In order to express the equation of motion in the form of a two-dimensional wave equation, we introduce the generalized tortoise coordinate:
\begin{align}
    \tilde{x} = 
    \frac{- 3}{\Lambda \sqrt{1+\zeta}}
    \bra{\frac{\tilde{r}_{-}\ln \norm{r-\tilde{r}_{-}}}{(\tilde{r}_{\rm c}-\tilde{r}_{-})(\tilde{r}_{\rm e}-\tilde{r}_{-})}
    +
    \frac{\tilde{r}_{\rm e}\ln \norm{r-\tilde{r}_{\rm e}}}{(\tilde{r}_{\rm e}-\tilde{r}_{\rm c})(\tilde{r}_{\rm e}-\tilde{r}_{-})}
    +
    \frac{\tilde{r}_{\rm c}\ln \norm{r-\tilde{r}_{\rm c}}}{(\tilde{r}_{\rm c}-\tilde{r}_{\rm e})(\tilde{r}_{\rm c}-\tilde{r}_{-})}
    },
\end{align}
up to an integration constant. 
We note that $\tilde{x} \to -\infty$ as $r \to \tilde{r}_{\rm e}$ and $\tilde{x} \to \infty$ as $r \to \tilde{r}_{\rm c}$. 
In terms of $\tilde{t}$ and $\tilde{x}$, the equation of motion is written as follows: 
\begin{align}
    \brb{
    \frac{\partial^{2}}{\partial \tilde{x}^{2}} - \frac{\partial^{2}}{\partial \tilde{t}^{2}} - V_{\ell}(\tilde{x})
    }
    \Psi_{\ell}
    = 0,
\end{align}
with 
\begin{align}
    V_{\ell}(\tilde{x}) = \frac{1}{1+\zeta} \bra{1 - \frac{r_{\rm g}}{r} - \frac{\Lambda_{\rm g}}{3}r^{2}} \brb{ \frac{\ell(\ell+1)}{r^{2}} - \frac{3 r_{\rm g}}{r^{3}} }.
\end{align}
As in the case of the stealth Schwarzschild solutions, we introduce the rescaled coordinates as follows: 
\begin{align}
    \tilde{T} = \frac{\tilde{t}}{\sqrt{1+\zeta}}, 
    \\
    \tilde{X}= \frac{\tilde{x}}{\sqrt{1+\zeta}}.
\end{align}
With these coordinates, the equation of motion can be rewritten as 
\begin{align}
    \brb{\frac{\partial^{2}}{\partial \tilde{X}^{2}} - \frac{\partial^{2}}{\partial \tilde{T}^{2}} - \tilde{V}_{\ell}(\tilde{X})}\Psi_{\ell}
    = 0,
    \label{eq:waveeqSdSsln}
\end{align}
with 
\begin{align}
    \tilde{V}_{\ell}(\tilde{X}) = \bra{1 - \frac{r_{\rm g}}{r} - \frac{\Lambda_{\rm g}}{3}r^{2}} \brb{\frac{\ell(\ell+1)}{r^{2}} - \frac{3r_{\rm g}}{r^{3}}}.
\end{align}
Therefore, even in the case of stealth Schwarzschild-dS solutions, we can recast the equation of motion into the standard Regge-Wheeler equation parametrized by $r_{\rm g}$ and $\Lambda_{\rm g}$. 
For the stealth Schwarzschild-dS solutions, the QNMs correspond to the modes that are purely ingoing at $r = \tilde{r}_{\rm e}$ and purely outgoing at $r = \tilde{r}_{\rm c}$. 
To determine the QNM frequencies, we consider the ansatz~$\Psi_{\ell} = \psi_{\ell}(\tilde{X})e^{-i \tilde{W} \tilde{T}}$. 
Let $\tilde{W}_{\ell, n}(r_{\rm g}, \Lambda_{\rm g})$ be the QNM frequencies obtained by solving Eq.~\eqref{eq:waveeqSdSsln} with the ansatz for $\Psi_{\ell}$ and
let $\omega^{\text{Sch-dS}}_{\ell, n}(r_{\rm s}, \Lambda)$ be the QNM frequencies calculated from the standard Regge-Wheeler equation parametrized by $r_{\rm s}$ and $\Lambda$. 
Then, the QNM frequencies $\tilde{W}_{\ell, n}(r_{\rm g}, \Lambda_{\rm g})$ can be expressed as 
\begin{align}
    \tilde{W}_{\ell, n}(r_{\rm g}, \Lambda_{\rm g}) = \omega^{\text{Sch-dS}}_{\ell,n} (r_{\rm g}, \Lambda_{\rm g}), 
\end{align}
where $\omega^{\text{Sch-dS}}_{\ell,n} (r_{\rm g}, \Lambda_{\rm g}) \coloneqq \omega^{\text{Sch-dS}}_{\ell,n} (r_{\rm s} \to r_{\rm g}, \Lambda \to \Lambda_{\rm g})$. 
Finally, from the relation between $\tilde{T}$ and $\tilde{t}$, the QNM frequencies of the odd-parity perturbations about the stealth Schwarzschild-dS solutions in the DHOST theory are given by
\begin{align}
    \omega^{\rm DHOST}_{\ell, n} 
    = 
    \frac{\omega^{\text{Sch-dS}}_{\ell,n} (r_{\rm g}, \Lambda_{\rm g})}{\sqrt{1+\zeta}}. 
    \label{eq:QNMDHOSTSdS}
\end{align}
That is, once we know the QNM frequencies in the Schwarzschild-dS spacetime parametrized by $r_{\rm g}$ and $\Lambda_{\rm g}$ in GR, we can find the QNM frequencies in the DHOST theory by applying the scaling law  Eq.~\eqref{eq:QNMDHOSTSdS}.

It is worth mentioning that there is a degeneracy between $r_{\rm s}$ and $\zeta$ as in the case of the stealth Schwarzschild solutions. 
As we discuss in Appendix~\ref{app:QNMsdsinGR}, the dimensionless QNM frequencies of the Schwarzschild-dS spacetime in GR, $\Omega^{\text{Sch-dS}}_{\ell,n} \coloneqq r_{\rm s}\, \omega^{\text{Sch-dS}}_{\ell, n}(r_{\rm s}, \Lambda)$, depends only on the combination~$r_{\rm s}^{2} \Lambda$, and hence we can write $\Omega^{\text{Sch-dS}}_{\ell,n} = \Omega^{\text{Sch-dS}}_{\ell,n} (r_{\rm s}^{2} \Lambda)$. 
Thus, the QNM frequencies of the Schwarzschild-dS spacetime can be written as 
\begin{align}
    \omega^{\text{Sch-dS}}_{\ell, n}(r_{\rm s}, \Lambda) = \frac{\Omega^{\text{Sch-dS}}_{\ell,n} (r_{\rm s}^{2} \Lambda)}{r_{\rm s}}.
\end{align}
As a result, Eq.~\eqref{eq:QNMDHOSTSdS} can be rewritten in terms of $\Omega^{\text{Sch-dS}}_{\ell, n}$ as follows:
\begin{align}
    \omega^{\rm DHOST}_{\ell, n} 
    = \frac{\Omega^{\text{Sch-dS}}_{\ell, n}(r_{\rm g}^{2} \Lambda_{\rm g})}{r_{\rm g} \sqrt{1+\zeta}}
    =
    \frac{\Omega^{\text{Sch-dS}}_{\ell, n} \bra{[r_{\rm s}(1+\zeta)^{3/2}]^{2} \Lambda}}{r_{\rm s} (1+\zeta)^{3/2}},
\end{align}
where we have used $r_{\rm g}=(1+\zeta)r_{\rm s}$ and $\Lambda_{\rm g} = (1+\zeta)\Lambda$. 
This means that, if we fix the value of $\Lambda$, $\omega^{\text{DHOST}}_{\ell,n}$ depends only on the combination~$r_{\rm s}(1+\zeta)^{3/2}$.
This shows that there is a degeneracy between $r_{\rm s}$ and $\zeta$ in the QNM frequencies for the stealth Schwarzschild-dS solutions as in the case of the stealth Schwarzschild solutions. 

For $r_{\rm g}^{2} \Lambda_{\rm g} \ll 1$, there is a perturbative formula for the QNM frequencies~\cite{Hatsuda:2023geo}. 
For $\ell=2$ fundamental mode, the formula is given by
\begin{align}
    \Omega^{\text{Sch-dS}}_{2,0}(r_{\rm g}^{2} \Lambda_{\rm g}) =  0.74734 - 0.17792\,i + \frac{9w_{1}}{2} r_{\rm g}^{2} \Lambda_{\rm g}  + \frac{81w_{2}}{8}r_{\rm g}^{4} \Lambda_{\rm g}^{2}  + \mathcal{O}(r_{\rm g}^{6} \Lambda_{\rm g}^{3}),
\end{align}
where $w_{1}= - 0.18649 + 0.03720\, i$, $w_{2}= - 0.04819 + 0.01428\, i$. 
Note that this formula implicitly assumes $r_{\rm s}\ne 0$ since otherwise the left-hand side cannot be defined.
Therefore, the QNM frequency~$\omega^{\rm DHOST}_{2, 0}$ becomes
\begin{align}
\begin{split}
        \omega^{\rm DHOST}_{2, 0} = \frac{1}{r_{\rm s} (1+\zeta)^{3/2}} \bigg( 0.74734 &- 0.17792\,i + \frac{9w_{1}}{2} \brb{r_{\rm s}(1+\zeta)^{3/2}}^{2} \Lambda
    \\
    &+ \frac{81w_{2}}{8}\brb{r_{\rm s}(1+\zeta)^{3/2}}^{4} \Lambda^{2}  + \mathcal{O}(r_{\rm s}^{6} \Lambda^{3})
    \bigg).
\end{split}
\end{align}
This explicitly shows that there is the degeneracy between $r_{\rm s}$ and $\zeta$ when we fix the value of $\Lambda$.

\section{Summary and discussions}
\label{sec:summary}

We have investigated the odd-parity perturbations about stealth Schwarzschild solutions and stealth Schwarzschild-de Sitter solutions with a linearly time-dependent scalar field in a subclass of DHOST theories, for which the deviation from general relativity is controlled by a single parameter~$\zeta$. 
We have derived the effective metric for the odd-parity perturbations and analyzed the characteristic curves. 
We have also shown that the Killing horizon(s) of the effective metric differs from that of the background metric. 
For $\zeta < 0$ case, the odd-parity perturbations can be superluminal and hence can escape from the region inside the Schwarzschild radius of the background metric, as demonstrated in Appendix~\ref{app:negativezeta}. 
We have derived the master equation for the odd-parity perturbations in the form of a two-dimensional wave equation, which can be expressed in the form of the standard Regge-Wheeler equation in GR with the rescaled black hole mass~$r_{\rm g}$ (and the rescaled cosmological constant~$\Lambda_{\rm g}$ in the case of the stealth Schwarzschild-de Sitter solutions). 
We have computed the QNM frequencies for both the stealth Schwarzschild solutions and the stealth Schwarzschild-dS solutions. 
In both cases, we have found that the QNM frequencies can be given by a simple scaling of those in GR.
In particular, we have shown that there is a degeneracy between the black hole mass~$r_{\rm s}$ and $\zeta$ in the QNM frequencies.

We have also solved an initial value problem for the odd-parity perturbations about the stealth Schwarzschild solutions employing the physically sensible formulation of the initial value problem proposed in~\cite{Nakashi:2022wdg}. 
We have defined a spacelike hypersurface~$\Sigma$ in the following manner: 
We have constructed a hypersurface~$\tilde{\Sigma}$ by slightly tilting the constant-$\tilde{t}$ surface. 
We have defined the region~$S$ where the surface~$\tilde{\Sigma}$ is spacelike. 
Furthermore, we have required that \eqref{req1}~the initial surface~$\Sigma$ coincides with $\tilde{\Sigma}$ in the region~$S$ within the numerical domain, and that \eqref{req2}~the initial conditions have a compact support in the region~$S$ within the numerical domain. 
We have analyzed the time evolution of a initial Gaussian wave packet. 
We have confirmed that the damped oscillation phase (ringdown phase) appears. 
We have found that a superposition of the QNMs in the DHOST theory is consistent with the numerical waveform through the fitting analysis. 
In particular, we have calculated the mismatch between the numerical waveform and the superposition of the QNMs in the DHOST theory and found that the minimum of the mismatch decreases and gets closer to the waveform peak when the overtones are taken into account. 
On the other hand, we have also confirmed that a superposition of the QNMs in GR does not well describe the numerical waveform. 
From these results, we conclude that the QNMs in the DHOST theory are excited in the physically sensible initial value problem.

We note that the perturbations about the stealth solution in DHOST theories would be strongly coupled~\cite{Babichev:2018uiw,deRham:2019gha,Motohashi:2019ymr,Takahashi:2021bml}.
A possible way out of this problem is to incorporate the scordatura term~\cite{Motohashi:2019ymr}.
However, as discussed in Sec.~\ref{subsec:odd}, we expect that the scordatura term would not lead to a qualitative change in our results on the odd-parity perturbations.
This is essentially because the strong coupling problem comes from the vanishing sound speed of the mode corresponding to the scalar degree of freedom, which belongs to the even-parity sector.
Along this line of thought, it would be intriguing to study the even-parity sector to see how the effect of the scordatura term shows up.
It should also be noted that the effect of modified gravity completely disappears in the odd-parity sector when $\zeta=0$, and hence the study of odd-parity perturbations alone cannot tell the difference from general relativity.
This is another motivation to study the even-parity sector.
We hope to come back to this issue in a future publication.

\section*{Acknowledgments}

This work was supported by JSPS KAKENHI Grant Nos.~JP22K03626 (M.K.), JP22K03639 (H.M.), JP22KJ1646 (K.T.), and JP23K13101 (K.T.) from the Japan Society for the Promotion of Science.

\appendix

\section{\texorpdfstring{Initial value problem for negative $\zeta$}{Initial value problem for negative zeta}}
\label{app:negativezeta}

In Sec.~\ref{subsec:initialvproblem}, we have analyzed the initial value problem for the odd-parity perturbations about the stealth Schwarzschild solutions in $\zeta>0$ case and shown that the waveform of the odd-parity perturbations is well described by a superposition of the QNMs in the DHOST theory. 
In this appendix, we analyze the case of $\zeta <0$. 
A remarkable property of the odd-parity perturbation for $\zeta<0$ is that the perturbation is superluminal in the whole spacetime. 
This implies that the odd-parity perturbations can escape from inside the Schwarzschild radius~$r_{\rm s}$.
As we mentioned in Sec.~\ref{subsec:tildetsch}, a constant-$\tilde{t}$ surface is always spacelike, and hence we choose it as the initial surface, i.e., we set $a=b=1$. 
Furthermore, we choose $\sigma$ and $\tilde{\mathcal{V}}_{0}$ so that the initial field profile has its support inside $r_{\rm s}$. 
\begin{figure}
    \centering
    \includegraphics[width=\textwidth]{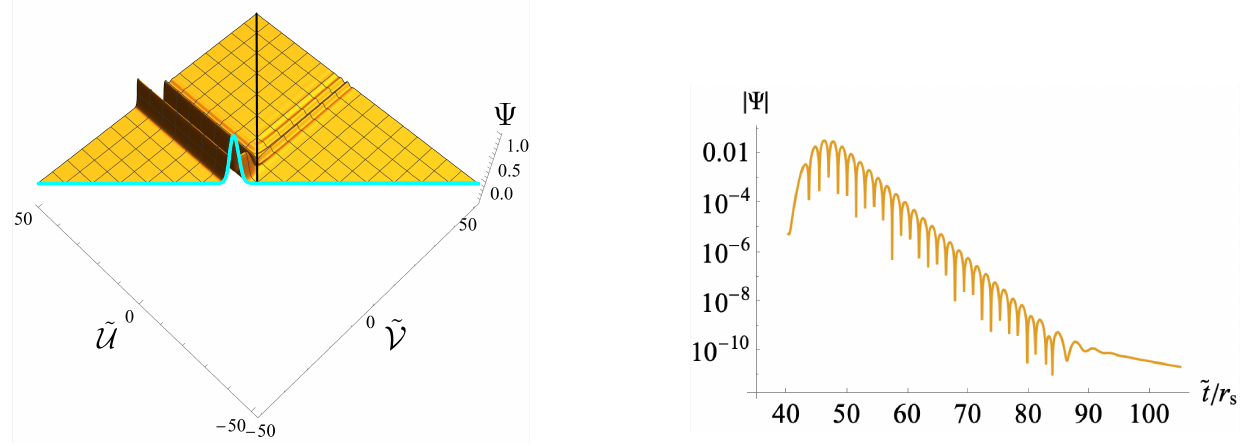}
    \caption{
    The evolution of the odd-parity perturbation in the $(\tilde{\mathcal{U}}, \tilde{\mathcal{V}})$-space (left panel) and the waveform of the perturbation observed by an observer at $\tilde{x}=40 r_{\rm s}$ (right panel) for $\zeta < 0$. 
    In the left panel, the cyan solid curve is the initial field profile and the black solid line is the location of the Schwarzschild radius~$r_{\rm s}$. 
    The left panel shows that the odd-parity perturbations can 
    escape from inside the Schwarzschild radius. 
    The right panel shows that the damped oscillation phase also shows up for $\zeta < 0$. 
    }
    \label{fig:negativezeta}
\end{figure}

Figure~\ref{fig:negativezeta} shows the evolution of the odd-parity perturbation for $\zeta = -0.5$ in the $(\tilde{\mathcal{U}}, \tilde{\mathcal{V}})$-space (left panel) and the waveform observed by an observer at $\tilde{x} = 40 r_{\rm s}$ (right panel). 
In the left panel, the cyan solid curve is the initial field profile, which has the compact support inside $r_{\rm s}$ depicted by the black solid line. 
The left panel explicitly shows that the odd-parity perturbation escapes from inside the Schwarzschild radius~$r_{\rm s}$.
The right panel shows that the damped oscillation phase also shows up in the $\zeta < 0$ case. 
We note that the damped oscillation phase can be well fitted by Eq.~\eqref{eq:fittingmodel} with $\mu = r_{\rm s}(1 + \zeta)^{-3/2}$.

\section{QNM frequencies of the Schwarzschild-dS spacetime in GR}
\label{app:QNMsdsinGR}

Here, we briefly mention a property of the QNMs of the Schwarzschild-dS spacetime in GR. 
The Regge-Wheeler equation for the Schwarzschild-dS solution in GR can be written as
\begin{align}
    \brb{
    \frac{\partial^{2}}{\partial x^{2}} - \frac{\partial^{2}}{\partial t^{2}} - V_{\ell}(x)
    }
    \Psi_{\ell} 
    = 0,
\end{align}
with the effective potential given by
\begin{align}
    V_{\ell}(x) = \bra{1 - \frac{r_{\rm s}}{r} - \frac{\Lambda}{3}r^{2}} \brb{ \frac{\ell(\ell+1)}{r^{2}} - \frac{3 r_{\rm s}}{r^{3}} }.
\end{align}
Substituting the ansatz~$\Psi_\ell=\psi_{\ell}(x) e^{-i \omega t}$, we have
\begin{align}
    \brb{
    \frac{\partial^{2}}{\partial x^{2}} +\omega^2 - V_{\ell}(x)
    }
    \psi_{\ell}(x) = 0.
\end{align}
In terms of the dimensionless coordinates~$\hat{r}\coloneqq r/r_{\rm s}$ and $\hat{x}\coloneqq x/r_{\rm s}$, the above equation takes the form
\begin{align}
    \brb{
    \frac{\partial^{2}}{\partial \hat{x}^{2}} + (r_{\rm s}\omega)^2 - r_{\rm s}^2V_{\ell}(x)
    }
    \psi_{\ell}(x) = 0.
\end{align}
Note that the effective potential written in terms of the dimensionless coordinates reads
    \begin{align}
    r_{\rm s}^2V_{\ell}(x)
    =\bra{1 - \frac{1}{\hat{r}} - \frac{r_{\rm s}^2\Lambda}{3}\hat{r}^{2}} \brb{ \frac{\ell(\ell+1)}{\hat{r}^{2}} - \frac{3}{\hat{r}^{3}} },
    \end{align}
where $r_{\rm s}$ and $\Lambda$ show up only in the combination~$r_{\rm s}^2\Lambda$.
As a result, the QNM frequencies normalized by $r_{\rm s}$ should depend only on $r_{\rm s}^2\Lambda$, which we write $\Omega^{\text{Sch-dS}}_{\ell, n}(r_{\rm s}^{2}\Lambda)$.
Therefore, the (dimensionful) QNM frequencies~$\omega^{\text{Sch-dS}}_{\ell, n}(r_{\rm s}, \Lambda)$ can be expressed as 
\begin{align}
    \omega^{\text{Sch-dS}}_{\ell, n}(r_{\rm s}, \Lambda) = \frac{\Omega^{\text{Sch-dS}}_{\ell, n}(r_{\rm s}^{2}\Lambda)}{r_{\rm s}},
    \label{eq:dimlessomega}
\end{align}

For $r_{\rm s}^{2} \Lambda \ll 1$, there is a perturbative formula for the QNM frequencies~\cite{Hatsuda:2023geo}. 
According to the formula, for $\ell=2$ fundamental mode, the dimensionless QNM frequency~$\Omega^{\text{Sch-dS}}_{2, 0}(r_{\rm s}^{2}\Lambda)$ is given by
\begin{align}
    \Omega^{\text{Sch-dS}}_{2, 0}(r_{\rm s}^{2}\Lambda) = 0.74734 - 0.17792\,i + \frac{9w_{1}}{2} r_{\rm s}^{2} \Lambda  + \frac{81w_{2}}{8}r_{\rm s}^{4} \Lambda^{2}  + \mathcal{O}(r_{\rm s}^{6} \Lambda^{3}),
\end{align}
where $w_{1}= - 0.18649 + 0.03720\, i$, $w_{2}= - 0.04819 + 0.01428\, i$. 
This is consistent with the fact that the dimensionless QNM frequencies $\Omega^{\text{Sch-dS}}_{\ell, n}$ depend only on $r_{\rm s}^{2} \Lambda$.

\bibliographystyle{JHEP}
\bibliography{bibliography}
\end{document}